\documentclass[11pt,a4paper]{article}

\usepackage{ascmac}
\usepackage{amsmath}
\usepackage{amssymb}
\usepackage{bm}
\usepackage[footnotesize,bf]{caption}
\usepackage{fullpage}
\usepackage{color}
\usepackage{float}
\usepackage{graphicx}
\usepackage[]{natbib}
\usepackage{subfigure}
\usepackage{txfonts}
\usepackage{threeparttable}
\usepackage{wrapfig}
\usepackage[]{multicol}
\usepackage{wrapfig}
\usepackage{authblk}

\usepackage{floatflt}

\title{\large Complete list of the ASTRO-H Science Working Group}
\date{\vspace{-0.5cm}}
\newcommand{\MakeWhitePaperTitle}{
	\begin{center}
		\begin{figure}
			\vspace{1cm}
			\begin{center}
				\includegraphics[width=0.2\hsize]{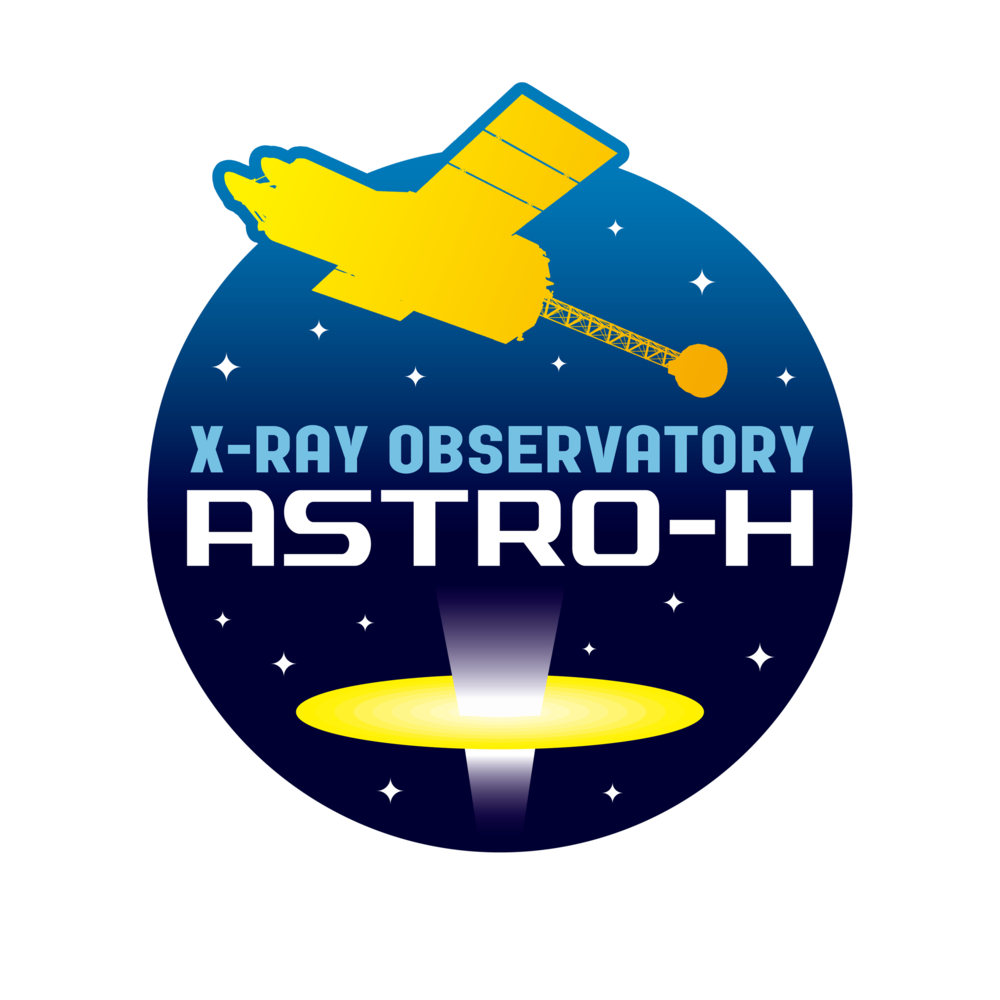}
			\end{center}
		\end{figure}
		\vspace{1cm}
		{\LARGE
		ASTRO-H Space X-ray Observatory\\
		White Paper\\
		}
		\vspace{5mm}
		{\large
		\WhitePaperTitle\\
		}
		\vspace{1cm}
		{
		\WhitePaperAuthors\\
		on behalf of the ASTRO-H Science Working Group
		}
	\end{center}
}

\usepackage{authblk}
\author[a]{Tadayuki~Takahashi}
\author[a]{Kazuhisa~Mitsuda}
\author[b]{Richard~Kelley}
\author[c]{Felix~Aharonian}
\author[d]{Hiroki~Akamatsu}
\author[e]{Fumie~Akimoto}
\author[f]{Steve~Allen}
\author[g]{Naohisa~Anabuki}
\author[b]{Lorella~Angelini}
\author[h]{Keith~Arnaud}
% ...
\affil[a]{Institute of Space and Astronautical Science (ISAS), Japan Aerospace Exploration Agency (JAXA), Kanagawa 252-5210, Japan}
\affil[b]{NASA/Goddard Space Flight Center, MD 20771, USA}
\affil[c]{Astronomy and Astrophysics Section, Dublin Institute for Advanced Studies, Dublin 2, Ireland}
\affil[d]{SRON Netherlands Institute for Space Research, Utrecht, The Netherlands}
\affil[e]{Department of Physics, Nagoya University, Aichi 338-8570, Japan}
\affil[f]{Kavli Institute for Particle Astrophysics and Cosmology, Stanford University, CA 94305, USA}
\affil[g]{Department of Earth and Space Science, Osaka University, Osaka 560-0043, Japan}
\affil[h]{Department of Astronomy, University of Maryland, MD 20742, USA}
% ...

\begin{document}

\newcommand{\WhitePaperTitle}{New Spectral Features}
\newcommand{\WhitePaperAuthors}{
	R.~K.~Smith~(SAO), H.~Odaka~(JAXA), M.~Audard~(Universit\`{e} de Gen\'{e}ve),
	G.~V.~Brown~(LLNL), M.~E.~Eckart~(NASA/GSFC), Y.~Ezoe~(Tokyo~Metropolitan~University),
	A.~Foster~(SAO), M.~Galeazzi~(University~of~Miami), K.~Hamaguchi~(UMBC/NASA),
	K.~Ishibashi~(Nagoya~University), K.~Ishikawa~(RIKEN), J.~Kaastra~(SRON),
	S.~Katsuda~(JAXA), M.~Leutenegger~(NASA/GSFC), E.~Miller~(MIT),
	I.~Mitsuishi~(Nagoya~University), H.~Nakajima~(Osaka~University), 
	T.~Ogawa~(Tokyo~Metropolitan~University),
	F.~Paerels~(Columbia~University),
	F.~S.~Porter~(NASA/GSFC), K.~Sakai~(JAXA), M.~Sawada~(Aoyama~Gakuin~University),
	Y.~Takei~(JAXA), Y.~Tanaka~(Kyoto~University), Y.~Tsuboi~(Chuo~University),
	H.~Uchida~(Kyoto~University), E.~Ursino~(University~of~Miami),
	S.~Watanabe~(JAXA), H.~Yamaguchi~(NASA/GSFC), and~N.~Yamasaki~(JAXA)
}
\MakeWhitePaperTitle

\newcommand{\astroh}{{\it ASTRO-H}}
\newcommand{\integral}{{\it INTEGRAL}}
\newcommand{\cgro}{{\it CGRO}}
\newcommand{\fermi}{{\it Fermi}}
\newcommand{\nustar}{{\it NuSTAR}}

\begin{abstract}
This white paper addresses selected new (to X-ray astronomy) physics and data analysis issues that will impact {\it ASTRO-H} SWG observations as a result of its high-spectral-resolution X-ray microcalorimeter, the focussing hard X-ray optics and corresponding detectors, and the low background soft $\gamma$-ray detector.  We concentrate on issues of atomic and nuclear physics, including basic bound-bound and bound-free transitions as well as scattering and radiative transfer.  The major topic categories include the physics of charge exchange, solar system X-ray sources, advanced spectral model, radiative transfer, and hard X-ray emission lines and sources.
\end{abstract}

\maketitle
\clearpage

\tableofcontents
\clearpage

\section{Introduction}

This white paper describes a number of new spectral features, arising from previously hard-to-observe processes, that will be detectable with {\it ASTRO-H}, primarily with the SXS detector but also using the hard X-ray and soft gamma ray detectors.  In particular, charge exchange (CX) between atoms and ions will create diagnostic soft X-ray emission lines that should be visible from a number of astrophysical environments, including solar wind interactions with planets, comets, and even the heliosphere.  In addition to CX with the solar wind CX, this may also occur in more distant regions such as supernova remnants and starburst galaxies, although this remains speculative.  Deep {\it ASTRO-H} observations with the SXS with non-dispersive high-resolution spectroscopy should reveal if CX is in fact occurring.  


\section{The Physics of Charge Exchange}

\begin{floatingfigure}[r]{4in}
\includegraphics[totalheight=3in]{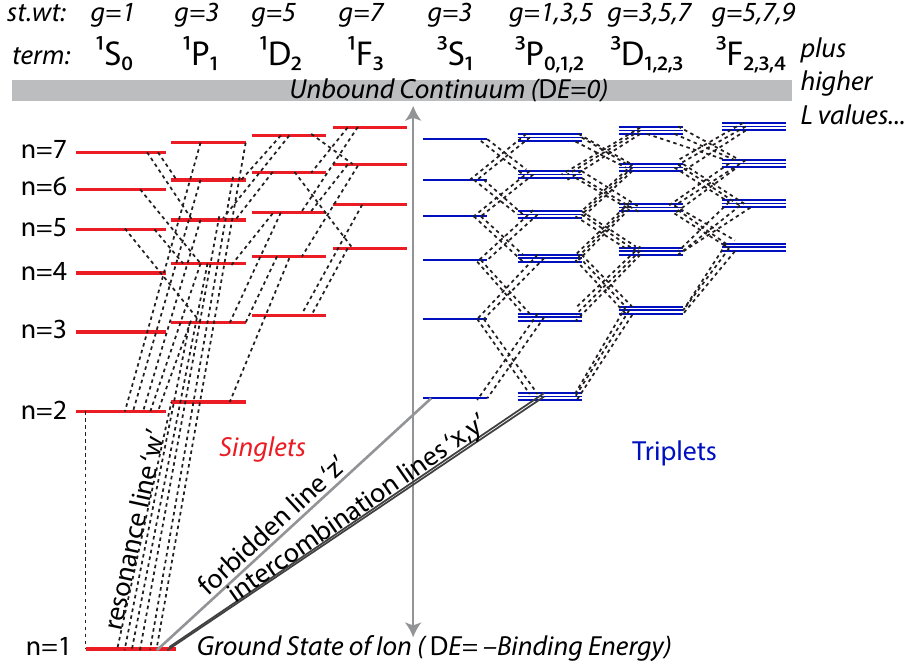}
\caption{Schematic Grotrian diagram for a Helium-like ion with selected radiative (downward) transitions marked as dashed lines.  Charge exchange transitions between a hydrogenic ion and a neutral hydrogen atom with binding energy -13.6 eV will result in an ionized hydrogen ion and a highly excited Helium-like state; for O VII this is $n \approx 6$.\label{fig:helike}
}
\end{floatingfigure}
The charge exchange (CX) process simply transfers an electron from one atom or ion to another.  Although there can be a change in momentum, no photon is created via the electron transfer itself.  Instead the electron maintains roughly the same binding energy in the process of moving from one atom to another, so the recipient ion is usually left in a highly-excited state that then radiatively stabilizes.  Typically, astrophysical charge exchange involves a donor hydrogen atom and a highly ionized metal recipient such as O$^{+7}$\ or Si$^{+10}$, although in some cases such as comets \citep{Cravens02} or planets in the Solar system \citep{Bhardwaj07} the donors can be metals or molecules.   Evaluating the impact of this process on X-ray spectra involves some fundamental difficulties, both in the atomic physics and the astrophysics.  

\vspace{5mm}
\subsection{Atomic Process}
From an atomic physics perspective, the exact cross section for charge exchange into a particular quantum state at low to moderate impact energies (relative to the binding energies) is difficult to calculate theoretically, although a number of results are now available for selected ions \citep[e.g.][]{Kharchenko01, Kharchenko03}.  Cross sections at larger energies are much easier to determine, and are in fact readily available from the fusion community (see, for example, \url{http://www.adas-fusion.eu/theme.php?id=2}).  CX provides an efficient means of heating tokamak plasmas since the fast-moving neutral atoms easily pass through the magnetic confinement and transfer their energy to the plasma.  

\begin{figure}
\includegraphics[totalheight=4.5in]{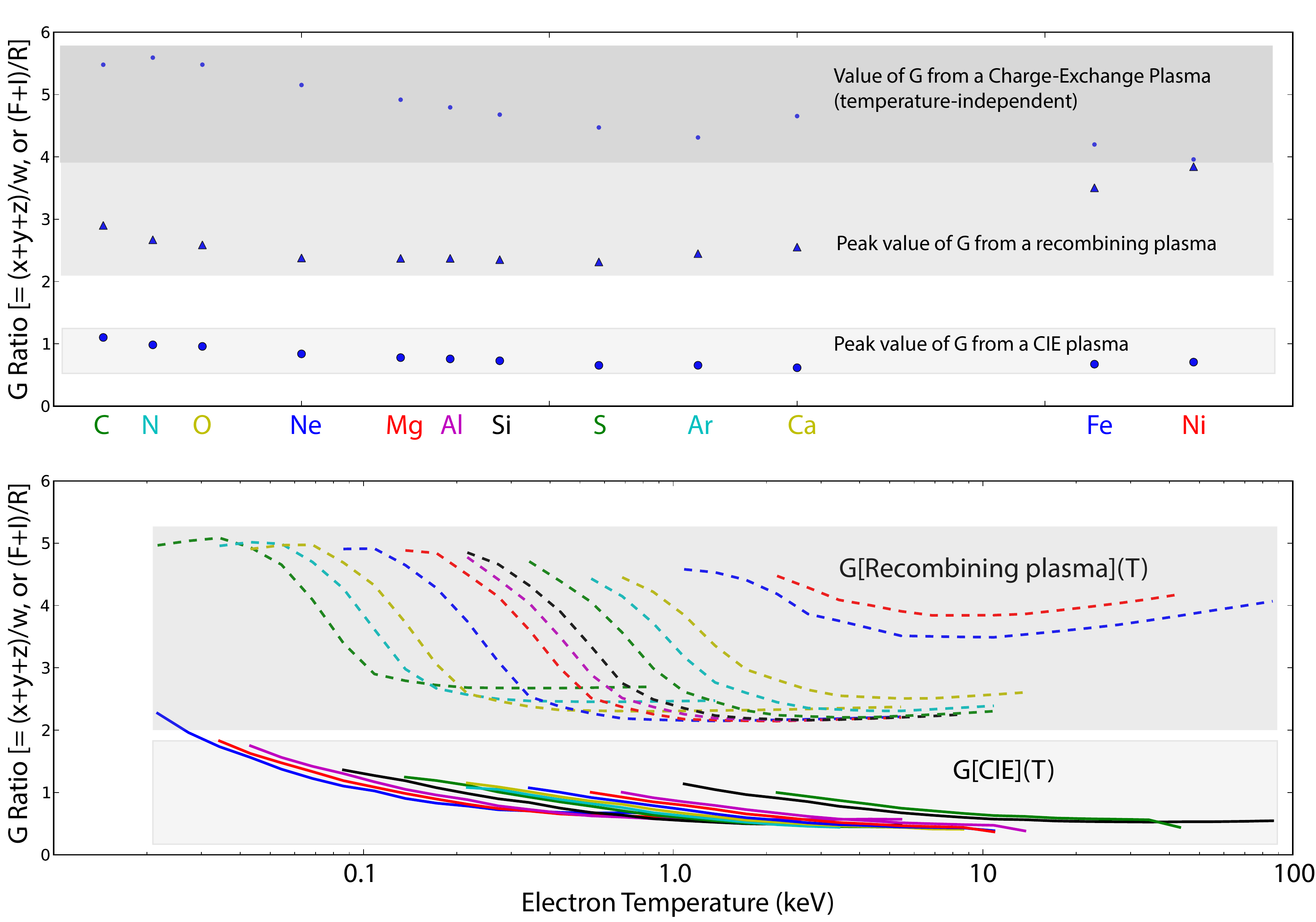}
\caption{[Top] The peak of the G-ratio from helium-like ions, defined as the ratio of the forbidden and intercombination lines to the resonance line, for each abundant element in three different common plasma situations: a CIE plasma, a recombining plasma, and a charge-exchange-dominated plasma.  [Bottom] The G ratio as a function of electron temperature in the CIE and recombining cases.  Note that CX is not shown as it is temperature-independent.\label{fig:Gratios}}
\end{figure}

At low energies, the simplest possible approximate cross section for CX is simply a constant, $\sigma_{\rm CX}\sim 3\times10^{-15}$\,cm$^{-2}$, which is normally accurate to about a factor of 3.  For a somewhat more accurate approximation one can follow \citet{Wegmann98}, who use a hydrogenic model for the CX cross section into the highly-excited state:
\begin{equation}
\sigma_{\rm CX} = 8.8\times10^{-17}{\rm cm}^2\ {{q-1}\over{ {{q^2}\over{2n^2}} - |I_p|}}
\end{equation}
where $q$\ is the charge of the ion, $I_p$\ is the ionization potential in atomic units (i.e., 1 au = 27.2 eV) of the target `donor' atom, and $n$\ the principal quantum number. Since the total cross section must be positive, $n$\ has a maximum value depending upon the charge of the ion.  For O$^{+8}$\ interacting with neutral H, therefore, $q = 8$, $I_p = 0.5$, and $n < 8$.  With $n=7$, the approximate cross section is $4\times10^{-15}$\,cm$^{-2}$, while for $n=6$\ it is $1.6\times10^{-15}$\,cm$^{-2}$.  Once the electron has been `exchanged' onto a highly-excited level of the ion, ``charge exchange'' photons resulting from radiative decays of the highly-exited ion are created (see Figure~\ref{fig:helike}).  Effectively, the electron decays towards the ground state like a pachinko ball, radiating photons at each step.  As implied by Figure~\ref{fig:helike}, this means that CX into a He-like triplet state will lead to enhanced forbidden and intercombination lines (as almost any exchange into a triplet state results in such lines).  It is worth noting that these lines can be created via normal collisional and photoionization processes as well; CX merely increases the rate of highly-excited transitions.  Thus CX into the He-like singlet state will lead to enhanced (relative to a purely collisional model in the ground state) high-$n$\ resonance transitions from $np {}^1P_1$\ to the ground state.  The atomic structure and radiative transition data can be used to determine the spectrum emitted as the ion stabilizes by radiative decay. In cases where the initial electron transfer would put the ion in a higher state than is available in AtomDB v2.0.2, which typically extends to $n=5$\ at a minimum and to $n=10$\ for the hydrogenic and helium-like isosequences, a new set of calculations (up to $n=14$) have been complete to enable the necessary spectral calculations. Figure~\ref{fig:Gratios} shows a series of comparisons of one common diagnostic, the He-like G ratio, for CX and other plasmas.  

A key residual uncertainty in low-energy CX cross sections is the angular momentum of the exchanged electron, which will impact the decay path and thus the emitted photons (e.g.\ in Figure~\ref{fig:helike}, the singlet versus triplet state).  The simplest possible approximation would be to simply distribute states evenly by the total angular momentum, or to distribute them by the statistical weight of each level.  \citet{Janev85} outlined more sophisticated theoretical models for state selective charge exchange (SSCX) models, depending on the relative velocity of the ions $v$ and the characteristic orbital velocity $v_\mathrm{o} = \sqrt{2I/m_e}$, where I is the ionization potential of the initial level of the donor electron (e.g. 13.6~eV for ground state H), and $m_e$ is the electron mass. For the case of the solar wind ($v\approx 400$~km~s$^{-1}$) and a hydrogenic donor, the low velocity regime ($v/v_\mathrm{o} < 1$) applies.  They considered two different distributions:
\begin{enumerate}
\item ``Landau-Zener:'', weighted by the function 
\begin{equation}
W(l)={{l(l+1)(2l+1)\times(n-1)! \times (n-2)!}\over{ (n+l)! \times (n-l-1)!}}
\end{equation}
\item ``Separable:'' weighted by the function 
\begin{equation}
W(l) = {{(2l+1)}\over{Z}}\times \exp\Big[{{-l \times(l+1)}\over{z}}\Big]
\end{equation}
\end{enumerate}
These two different distributions do not lead to large changes in the resulting spectrum, so both should be considered as possibilities that partially address the range of uncertainties in this approach.

\subsection{Astrophysical Modeling}
\begin{floatingfigure}[r]{3.5in}
\includegraphics[totalheight=2.5in]{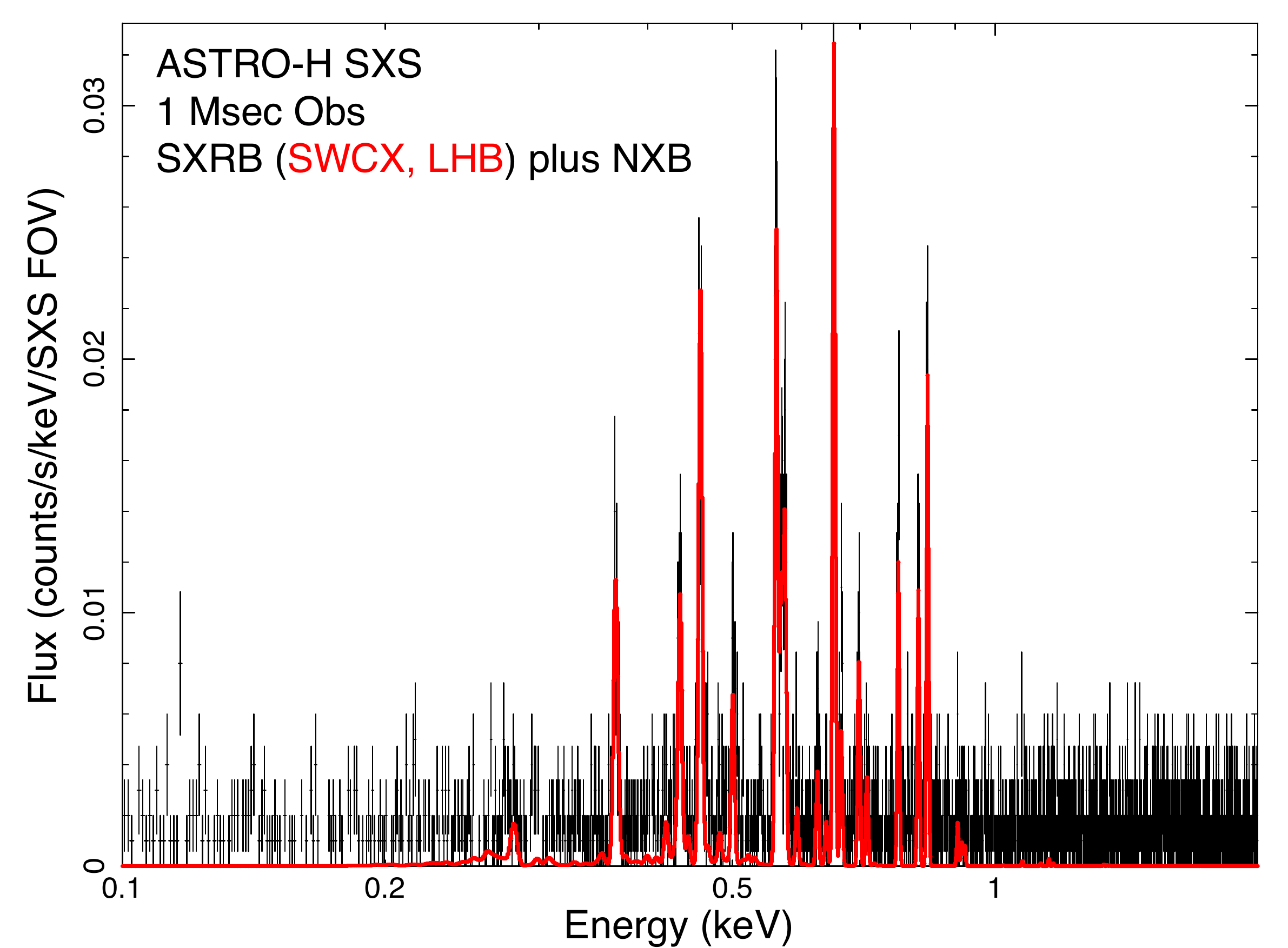}
\caption{Soft X-ray background model, simulated using simx, assuming a combined LHB and SWCX model that fits the DXS spectrum and CHIPS upper limits (Smith et al. 2014). \label{fig:sxs-sxrb}}
\end{floatingfigure}
A more significant problem exists in the astrophysical modeling, as it is not possible to determine with any accuracy the density of donor atoms and recipient ions in a given region of space.  Individual situations do exist where these can be estimated, such as charge exchange in the solar wind  where ions in the solar wind are directly measured by satellites like ACE and WIND and models exist for the geocoronal or heliospheric neutral atoms \citep{Koutroumpa06,Koutroumpa09}.   Thus, while there remain many ions without explicit cross section calculations, these are matched by our lack of knowledge of the mixing in hot plasma / dense cloud interaction regions that could lead to CX spectra. Although not ideal, these difficulties mean that any approximations in the model will not dominate the inherent uncertainties in the final result.

The CX rate along a line of sight during an astrophysical observation can be written as $F_{CX} = (4\pi D^2)^{-1} \int{n_D n_R v \sigma^{cx}_{D\rightarrow R}(E) dV}$\ where $N_{CX}$ is the number of CX reactions in the volume, $n_D$ and $n_R$ are the donor and recombining ion densities, and $\sigma^{cx}_{D\rightarrow R}(E)$ is the energy-dependent cross section for charge exchange between the donor and recombining ions at velocity $v$. In the case of solar wind charge exchange (SWCX), the donor ions will be neutral hydrogen or helium atoms in the interstellar medium, while the recombining ions will be the solar wind ions. Of the  parameters determining the CX rate, both $n_R$ and $n_D$, and their variation within the volume are unknown, and $v$\ can only be roughly estimated. In the interest of developing an approximate model, fine accuracy for the total charge exchange cross section can be set aside in order to obtain relative partial capture cross sections into different $n$ and $l$ shells, as described above.  This can be used to predict the shape of the CX spectrum, although absolute magnitudes will have to be obtained from fits to spectra and discretion during analysis.

The actual line strengths in this approximation can be determined using the atomic structure and radiative transition data available in the AtomDB v2.0.2 (\url{http://www.atomdb.org}). If the initial electron transfer would put the ion in a higher state than is available in the AtomDB (which typically extends to $n=5$\ at a minimum for ions that can emit X-rays in a collisional plasma and to $n=10$\ for the hydrogenic and helium-like isosequences), then more atomic data are needed.  For these ions \citet{Smith12, Foster13} performed a large atomic structure calculation to determine the energies and radiative transition rates of all levels up to at least $n'$, the predicted state of the post-CX system. We calculated all ions of astrophysical interest up to $n� \le 13$\ (in Ni$^{+27}$) using the \textsc{Autostructure} \citep{badnell1986} code. The largest number of levels needed (for the ion Ni$^{+20}$) was 2,374. Until fairly recently, these calculations were simply not feasible given computational limitations, but now the primary limitation is the time required to set up the calculations and to check the results.

A first cut of this spectral model that can be used within XSPEC (or ISIS) is available at \url{http://www.atomdb.org/acx}.  \citet{Smith14} used this model to fit the Diffuse X-ray Spectrometer (DXS) data as well as the Cosmic Hot Interstellar Plasma Spectrometer (CHIPS) using a model that combines a hot thermal plasma (representing the Local Hot Bubble) plus a two component SWCX model (representing the Fast and Slow solar winds).  While there are strong indications that the SWCX components are time-variable, this should give a reasonable idea of what level of line emission could be expected. A simulated 1 Msec spectrum is shown in Figure~\ref{fig:sxs-sxrb} for the SXS, including the non-X-ray-background (NXB) as well.

\vspace{5mm}
\section{Charge Exchange in the Solar System and Beyond}

In recent years, many sources of soft X-rays within the solar system, besides the Sun itself, have been discovered. This has opened the door to new studies of the mechanisms and the physics responsible for the emission.

In particular, in Solar Wind Charge eXchange (SWCX), highly ionized ions in the solar wind collide with neutral atoms picking up an electron in an excited state. The electron then decays producing UV and X-ray emission. SWCX was originally discovered from observations of comets, where solar wind would ``charge exchange'' with the neutrals in the comet \citep{cra97}. It is now believed that solar wind also ``charge exchanges'' with neutrals in our atmosphere (magnetosheath SWCX) (Cravens 2001, Robertson and Cravens 2003a,b) and with interstellar neutrals in the solar system (heliospheric SWCX) \citep{Cox98}.

Many solar system planets have also been recognized to show X-ray emission. The detected objects are Mercury, Venus, Earth, Mars, Jupiter, and Saturn (see \citealt{bha07,ezo11a,ezo12} for review).

Preliminary studies with current and past missions have been quite fruitful in laying the groundwork for the understanding of the X-ray emission and the mechanisms behind them. However, this studies are still in their infancy and many of the fundamental physical parameters are still unknown. In particular, current missions lack the energy resolution necessary to resolve the many emission lines associated with these processes and any significant response in the $1/4$~keV energy band, where a significant fraction of the emission lies. The time required for each observation and the time variation of the emission has also limited the number of observations available. 

The high energy resolution of the {\it ASTRO-H} SXS instrument should bring significant improvements in at least two if these areas.


\subsection{Introduction} 

Fully characterizing and modeling the SWCX emission requires an understanding 
of the distribution of neutral material in the magnetosheath and heliosphere, 
the properties and distribution of the solar wind, and the interaction 
cross-sections. SWCX emission probes both the solar wind and the distribution 
of neutral atoms throughout the solar system, complementing many other 
heliophysical measurements. 

However, understanding SWCX emission is also critical to astrophysical 
observations, as it contributes a significant background to X-ray observations 
of extended objects that may be inseparable from the signal of interest, 
as many of the emission lines are the same used as plasma diagnostics of thermal 
emission. It is, in particular, essential to understanding the nature and 
structure of the Local Hot Bubble (LHB) surrounding the Sun.  Finally, soft 
X-ray studies of planets also offer a new tool for remotely studying their 
atmosphere and environment. 

\smallskip
\noindent {\bf Solar Wind Charge eXchange}

Four components of SWCX can be identified and characterized using {\it ASTRO-H}:
\begin{enumerate}
\item Heliospheric
\item Geocoronal
\item Cometary
\item Planetary
\end{enumerate}

The first two components directly affect the emission from the Diffuse 
X-ray Background (DXB) and other extended object emission. The last two can be 
used directly to probe the physics and composition of the objects themselves.
Cometary SWCX can also be used to study the properties of 
``fast'' solar wind, which is emitted at high ecliptic latitude.
SWCX is the only component of the DXB with a ``short scale'' temporal variation 
due to the flux of ions in the solar wind that can be used to identify it on top 
of the other components. 

Geocoronal, planetary, and cometary SWCX are varying on a very short time scale 
of hours to days, due to the interaction of solar wind with neutral concentrated 
in a small volume (a planet's atmosphere or a comet surface), while heliospheric 
SWCX varies more smoothly on a time scale of several days to months, due to 
the diffusion time of the wind through the solar system.  An annual modulation 
due to distributions of interstellar neutrals (H and He) passing through the solar 
system is also present.

Above 0.3 keV SWCX emission consists mostly of K-shell lines of C, N, O, Ne, and Mg, 
and L-shell lines of Fe. To first order, the line intensity depends linearly on 
solar wind conditions. In particular, the intensity $I$  is the integral over the 
line of sight $ds$ of the product of the ion density $n_{ION}$, the neutral density 
($n_H$ and $n_He$), and the interaction cross-section 
\begin{equation}   
I_{O VII}=\int \alpha_{ION} n_{ION}  (n_H+n_{He}) \langle g \rangle ds,
\end{equation}
where $\langle g \rangle$ is average solar wind speed, and the cross-sections $\alpha_{ION}$ 
are velocity dependent. 

Solar wind data are available through the ACE, WIND, and STEREO-A/B missions.  
The combination of SWEPAM and SWICS instruments aboard ACE provide $\langle g \rangle$ and $n_{ION}$ 
(at least for oxygen), while an  {\it ASTRO-H} campaign will be essential to measure the cross 
sections $\alpha_{ION}$ for different solar wind conditions (as they are velocity 
dependent), the distribution of neutrals $n_H$ and $n_{He}$ as a function of pointing 
direction, and the ratio n$_{\rm OVIII}$/n$_{\rm OVII}$\ as a function of solar wind 
conditions. In addition to the invaluable contribution to the physics of SWCX, such 
parameters are paramount for the future prediction and evaluation of SWCX contribution 
to astrophysical observations with existing ({\it XMM-Newton}, {\it Chandra}, {\it Suzaku}) observations 
and future missions (eROSITA, AXSIO, Athena).

Directly evaluating the effect of background due to SWCX on observations of diffuse 
object with current and future missions is, in general, impossible, and modeling 
its effect remains, in many cases, the only long-term viable option. While there 
are well-defined models for SWCX emission from the heliosphere, testing them is 
problematic. Koutroumpa et al. (2007), using a self consistent model of the heliospheric 
SWCX emission, managed to associate observed discrepancies in {\it XMM-Newton} and {\it Suzaku}
observations separated by several years with solar cycle-scale variations of the 
heliospheric SWCX emission. A more recent paper (Snowden 2009) was successful in 
using an {\it XMM-Newton} observation to test a model (Robertson \& Cravens 2003a) for 
emission from the near-Earth environment. 

Koutroumpa et al. (2009) uses a different series of {\it XMM-Newton} observations to 
search for SWCX emission from the He focusing cone. They were relatively successful, 
but the detection was at a marginal level due to the short exposures and higher 
backgrounds experienced by {\it XMM-Newton}. A critical aspect of all future investigations 
of extended objects will be to test and upgrade our current SWCX models, specifically 
where it pertains the input parameters discussed above. 

\smallskip
\noindent {\bf SWCX, Local Hot Bubble, and extended sources}

It has been suggested that SWCX emission may actually be responsible for all 
of the ``local flux'' observed at $1/4$ keV near the galactic plane (Lallement 2004a,b) 
and that much of the observed $3/4$ keV emission may also originate from SWCX (Koutroumpa 2007), 
leaving little room for LHB emission. It is unclear whether roughly half of the 
observed background originates within the nearest 100 AU or from a region that extends 
to several hundred parsecs, a dynamic range, and therefore uncertainty, of five orders 
of magnitude in the location of the source. Whether the observed emission originates 
in the heliosphere or the LHB has a considerable impact on our understanding of the 
local interstellar medium and the local energy balance.

Plasma emission and SWCX emission consist of similar lines, making them difficult to 
separate with current instruments. However, the process responsible for the line 
excitation is quite different, making the relative ratio between lines (e.g., within the
O VII triplet) significantly different. The {\it ASTRO-H} SXS instrument would be the first 
instrument with sufficient area and sufficient observing time that could be used to 
clearly identify individual lines and their ratio. With an accurate separation between 
LHB and SWCX emission we can determine the distribution of the hot plasma within the 
LHB. Combining this information with the geometry of the local cavity derived from 
other wavelengths, typically interstellar absorption line measurements, we can then 
derive the physical parameters of the plasma, that is, the pressure and density. 
Knowing the physical conditions of the plasma will lead to more accurate pictures 
of the solar neighborhood and the evolution of bubbles of hot gas produced by 
supernovae or stellar wind when they near the end of their existence.

The contribution from SWCX significantly affect also our understanding of the physical 
parameters of the intra-group medium and the halo of the Milky Way which concerns the 
energy balance of the Galaxy as a whole. SWCX likely causes the short-term temporal 
variations in the X-ray background that complicate low surface brightness observations 
with {\it Suzaku}, {\it XMM-Newton}, and {\it Chandra}. Nearly every low surface brightness observation 
has some contamination from this foreground emission. 

In addition to the DXB, contamination by SWCX impacts the accuracy of observations of 
low surface brightness objects that fill the field of view of X-ray observatories, such 
as  hot plasmas in nearby superbubbles, the interstellar medium in our galaxy, the 
Magellenic clouds, nearby galaxies, and even clusters of galaxies. Current analysis 
procedures remove time intervals showing variation, however, this does not identify 
the slowly changing heliospheric SWCX emission and results in a misleading interpretation 
of the spectral content of the source (e.g., Henley 2008). These emission regions 
subtend large solid angles relative to the current instrumentation so observations 
cannot be ``cleaned'' using data from the observation itself. Using a separate 
pointing to subtract the background is also unfeasible due to the SWCX time variability. 
As a result, without a clear understanding of the SWCX background there will be 
complete uncertainty in interpreting some observations and there will always be some 
level of uncertainty in the interpretation of all observations of astrophysical plasmas 
and this will be the case for all observatories. It is essential to construct effective 
models of the foreground SWCX emission and validate these models experimentally, 
otherwise the absolute calibration of all major X-ray observatories for soft, 
low-surface brightness objects will be uncertain and spectral modeling will lead to 
incorrect physical interpretation of sources.

\smallskip
\noindent {\bf X-ray emission from planets}

X-ray emission mechanisms of the planets can be roughly divided into three classes. 
The first class is due to scattering of solar X-rays from neutrals. This needs a large 
neutral column density of $>$10$^{19}$ cm$^{-2}$ that is easy to be obtained in a planetary 
upper and lower atmosphere but difficult in a more tenuous exosphere. This type is often 
called disk emission. 
The second class is a charge exchange (CX) of solar wind ions (keV/amu) or energetic 
ions (MeV/amu) accelerated in planetary magnetospheres with neutrals. Due to its large 
cross sections at solar wind ion energies, typically a few times 10$^{-15}$ cm$^2$, this 
type of emission can be seen in the planetary exosphere and is sometimes called halo emission. 
The third class is continuum and line emission by high energy electrons accelerated 
in planetary magnetic fields. This class has been seen in the vicinity of Earth and Jupiter.

\subsection{Heliospheric SWCX}

Different approaches have been used to separate heliospheric SWCX from other components 
of the diffuse X-ray background. In particular, its temporal variation, spatial variation, 
and line ratio have all been suggested and investigated. The use of {\it ASTRO-H} SXS instrument 
is unique in that it can combine both temporal variation and spectroscopic information 
to separate the SWCX emission from the other DXB components and study its structure and 
emission parameters. 

The best long term approach to achieve this goal is through a monitoring 
campaign, however, significant, breakthrough science, can be achieved with a single dedicated
pointing.


\subsection{Planets} 

\noindent {\bf Background and Previous Studies:}

\begin{figure}
\begin{center}
\includegraphics[width=0.8\hsize]{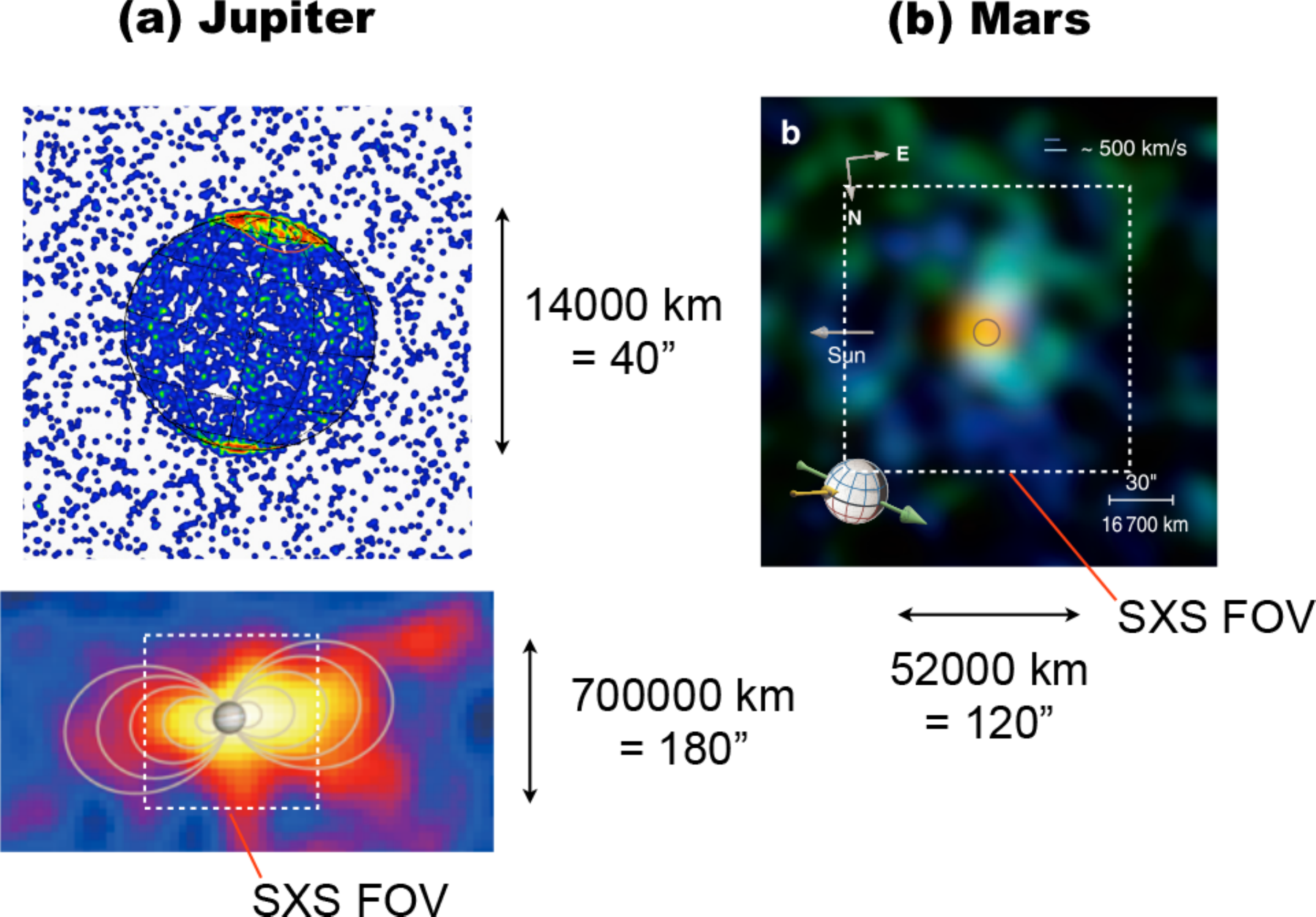}
\caption{
(a) X-ray images of Jupiter with {\it Chandra} HRC in 0.5--10 keV \citep{gla02} and
Jupiter's inner radiation belts with {\it Suzaku} XIS in 1--5 keV \citep{ezo10a}. In the bottom panel, 
solid lines indicate equatorial crossings of Jovian magnetic fields. An optical image of 
Jupiter is overlaid.
(b) Soft X-ray image of Mars with {\it XMM-Newton} RGS \citep{den06}. Colors indicate ionized 
oxygen emission lines in blue, ionized carbon lines in green, and fluorescence lines in yellow and red.
}
\label{fig:jupiter-mars-ximage}
\end{center}
\end{figure}

In the last decade, our knowledge about the X-ray emission from planets has been greatly advanced 
by the X-ray astronomy satellites {\it Chandra}, {\it XMM-Newton}, and {\it Suzaku}. 
Figure \ref{fig:jupiter-mars-ximage} (a) shows X-ray images of Jupiter. 
Jupiter is the most luminous planet ($\sim$1-2 GW) in X-rays. Its X-ray emission is composed of 
auroral emission due to CX and keV electron bremsstrahlung and disk emission due to scattering 
of solar X-rays \citep{bra04}.
The auroral ions responsible for CX could have energies of several MeV/amu.
The origin of the ions has been matter of debate and there are two kinds 
of possibilities: high charge state energetic ions 
precipitating directly from solar wind, or magnetospheric ions from Io's 
volcano, both of which can be accelerated by strong field potential in 
Jupiter's polar regions (e.g., \citealt{cra03}). 
{\it Suzaku} discovered diffuse X-ray emission 
associated with Jupiter's inner radiation belts, which possibly
arising from inverse Compton scattering of solar photons by electrons with energies of tens of MeVs \citep{ezo10a}.
In-situ measurements and ground based observations have shown the existence of 
Jupiter's high energy Van Allen radiation belts for electrons up to 50 MeV.
Hence, if this emission is truly inverse Compton scattering, hard X-ray observations 
can constrain the population and spatial distribution of tens MeV electrons.
In fact, using the X-ray morphology and luminosity, \cite{ezo10a} estimated a necessary 
electron density and found that it exceeds an empirical model based on past in-situ 
measurements by a factor of at least 7, although the exact reason for this discrepancy 
is unclear. 

Figure \ref{fig:jupiter-mars-ximage} (b) shows an X-ray image of Mars, whose X-ray luminosity
can reach $\sim$130 MW at maximum. 
The emission originates not only from the disk but also outflowing components from the poles to 
the antisolar direction, most probably caused by CX between solar wind ions and neutrals in 
the Martian exosphere \citep{den06}. 

In spite of these discoveries, limited energy resolution and insufficient sensitivity to X-rays above 10 keV 
leave fundamental problems such as:

\begin{description}
\item[Jupiter's aurora] :
How and to what energy are ions and electrons accelerated in the magnetosphere~? 
\item[Jupiter's inner radiation belts] : 
Can hard X-ray observations be used to constrain the spectral and spatial distributions 
of $\sim$10 MeV electrons~? 
\item[Martian exosphere] : 
How and what amount of the Martian exosphere is escaping~? 
Can soft X-ray observations be used to constrain the atmospheric escape~?
\end{description}

\smallskip
\noindent
{\bf Prospects \& Strategy:}

The {\it ASTRO-H} SXS opens up a new view of the planetary environments using the CX emission lines.
We can directly distinguish fluorescent emission lines from the CX lines such as OVII K$\alpha$ in 
the Jupiter's aurora and Martian exosphere, and can accurately determine line properties.
Although grating observations have been conducted with {\it XMM-Newton} RGS and provided hints 
of what high energy resolution spectroscopy could achieve, the spatial extent of all solar system planets combined with the limited effective area of the RGS leads to significant uncertainties in the line properties. 
The SXS can achieve high energy resolution and high S/N at the same time because of its large 
effective area above 0.5 keV and anticipated low background.

Simultaneous observations with exploration satellites (JUNO, Mars Express) 
and solar wind monitoring satellites (ACE, WIND) will enable us to deduce the exospheric 
density from the observed X-ray line intensities, as demonstrated with {\it Suzaku} for Mars 
\citep{ish11}. This will lead to an understanding of the not-well known exospheres.  

From Doppler shift and broadening, dynamics of ions in the line of sight direction will also be
constrained with SXS, since there are significant variations expected. 
In particular, the average speed of the solar wind particles ranges from $\sim$400 km to $\sim$700~km~s$^{-1}$, 
while that of the precipitating ions accelerated by strong field aligned potential into Jupiter's aurora is presumed as $\sim$5000~km~s$^{-1}$ \citep{bra07}.

The {\it ASTRO-H} SXI and HXI will allow us to detect hard X-ray continuum due to the inverse Compton 
scattering at the inner radiation belts and bremsstrahlung at the aurora. Hence, not only 
high energy ions but also electrons can be investigated from high sensitivity X-ray spectra.

\smallskip
\noindent
{\bf Beyond Feasibility:}

If the solar wind ion flux dramatically increases during a rare occasion such as a coronal 
mass ejection, the planetary CX emission can significantly increase. A TOO-type joint 
observation with solar wind monitoring satellites may make this rare detection feasible.


\subsection{Geocoronal SWCX}

\noindent
{\bf Background and Previous Studies:}

\begin{floatingfigure}[r]{4in}
\includegraphics[width=90mm]{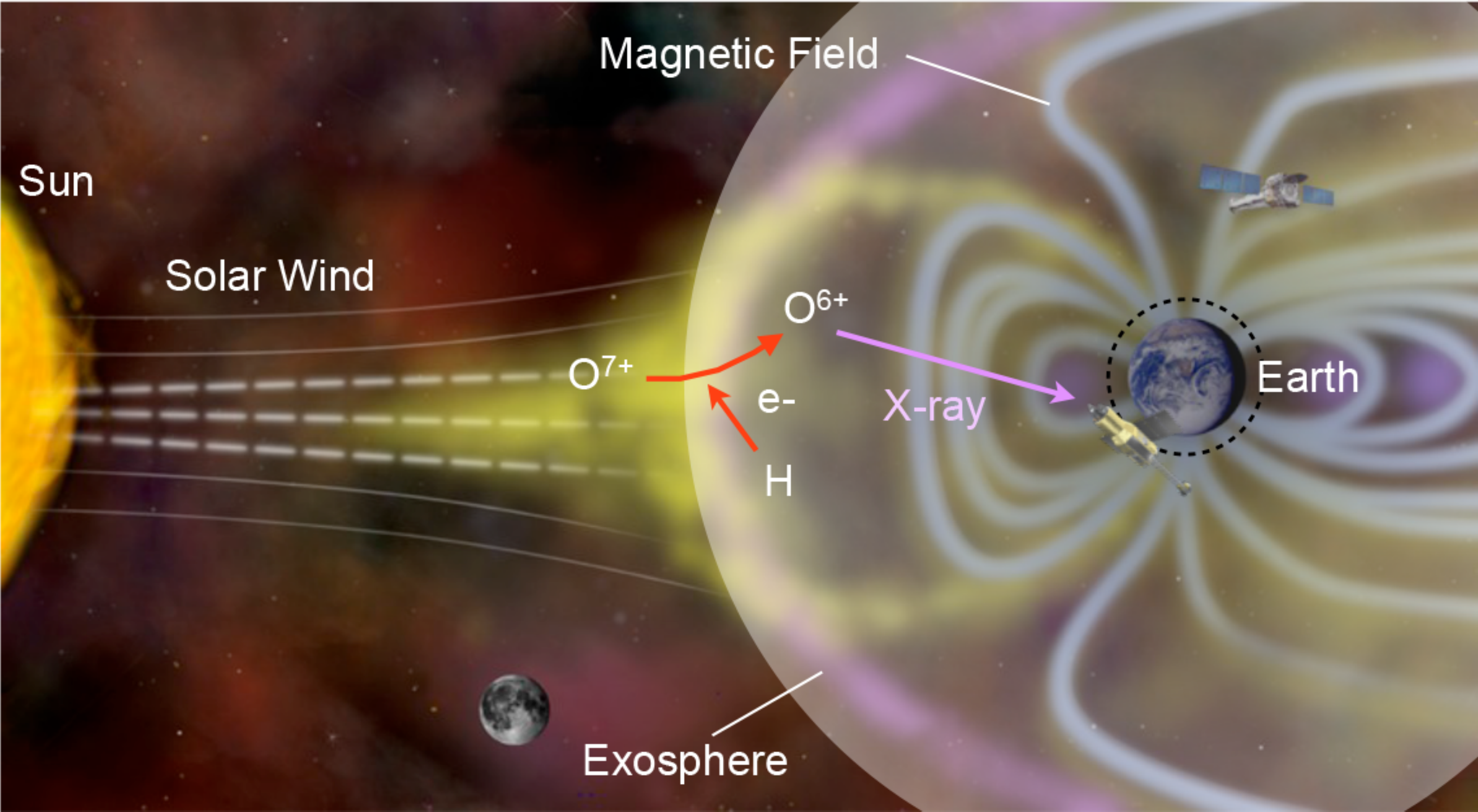}
\caption{Schematic of geocoronal SWCX emission modeled after an illustration taken from 
http://chandra.harvard.edu/photo/2003/moon. \label{fig:geocorona-concept}}
\end{floatingfigure}
Solar wind charge exchange (SWCX) in the Earth's exosphere, which is often called geocoronal
SWCX, can constitute additional sources of diffuse time-variable background in astrophysical observations. 
Figure \ref{fig:geocorona-concept} shows a schematic of the process.
The SWCX emission is in general composed of emission lines from highly ionized ions and 
mainly seen in the soft X-ray band ($<2$ keV).
This emission was found during the ROSAT all sky survey as ``Long Term Enhancements'' \citep{sno94}.
A systematic search for the SWCX events has been done for the {\it XMM-Newton} and {\it Suzaku} archival data
by \cite{car11} and \cite{ish13}, respectively.
This unwanted background for general users can be a new tool 
to study the solar wind as well as the Earth's exosphere and magnetosphere as demonstrated with
{\it Suzaku} observations (\citealt{fuj07,ezo10b,ezo11b}).
However, the high energy resolution spectroscopy has been impossible because the geocoronal 
SWCX is extended over the FOV of X-ray CCDs.

\smallskip
\noindent
{\bf Prospects \& Strategy:}

For the first time with the {\it ASTRO-H} SXS, the high energy resolution spectroscopy will
be possible for the geocoronal SWCX. 
In all SXS observations of astronomical objects, we can constrain the SWCX contribution 
directly from line distributions, since the CX is known to produce strong emission lines 
from large principle quantum number states and strong forbidden lines. 
The geocoronal SWCX can be separated from another SWCX background, i.e., the heliospheric 
SWCX using short term ($<$1 day) time variation because of its compact emission region 
($\lesssim$tens Earth radii).

Combined with ACE and WIND data, the the solar wind ion fluxes can be obtained 
at the same time. 
The observed X-ray intensity will thus be able to be compared with a theoretical model
considering an exospheric density model that gives the neutral distribution and a 
geomagnetic field model that defines the solar wind distribution.
As suggested from {\it Suzaku} observations (e.g., \citealt{fuj07}), this comparison will 
provide us a clue of the shape of the magnetosphere, especially the location of the 
magnetocusp, which is recognized as one of the most important issues in space plasma physics.

From Doppler shift and broadening of individual emission lines detected in spectra, dynamics 
of ions will be constrained with an accuracy of several hundreds km~s$^{-1}$. Solar wind abundances 
will be studied from the relative X-ray line intensities. 
Furthermore, since the geocoronal SWCX is the nearest CX source, it will provide us with 
fundamental information on CX such as line ratios and cross sections, that must be useful 
for extra-solar observations.

\smallskip
\noindent
{\bf Targets \& Feasibility:}

\begin{floatingfigure}[r]{8.8cm}
\includegraphics[width=85mm]{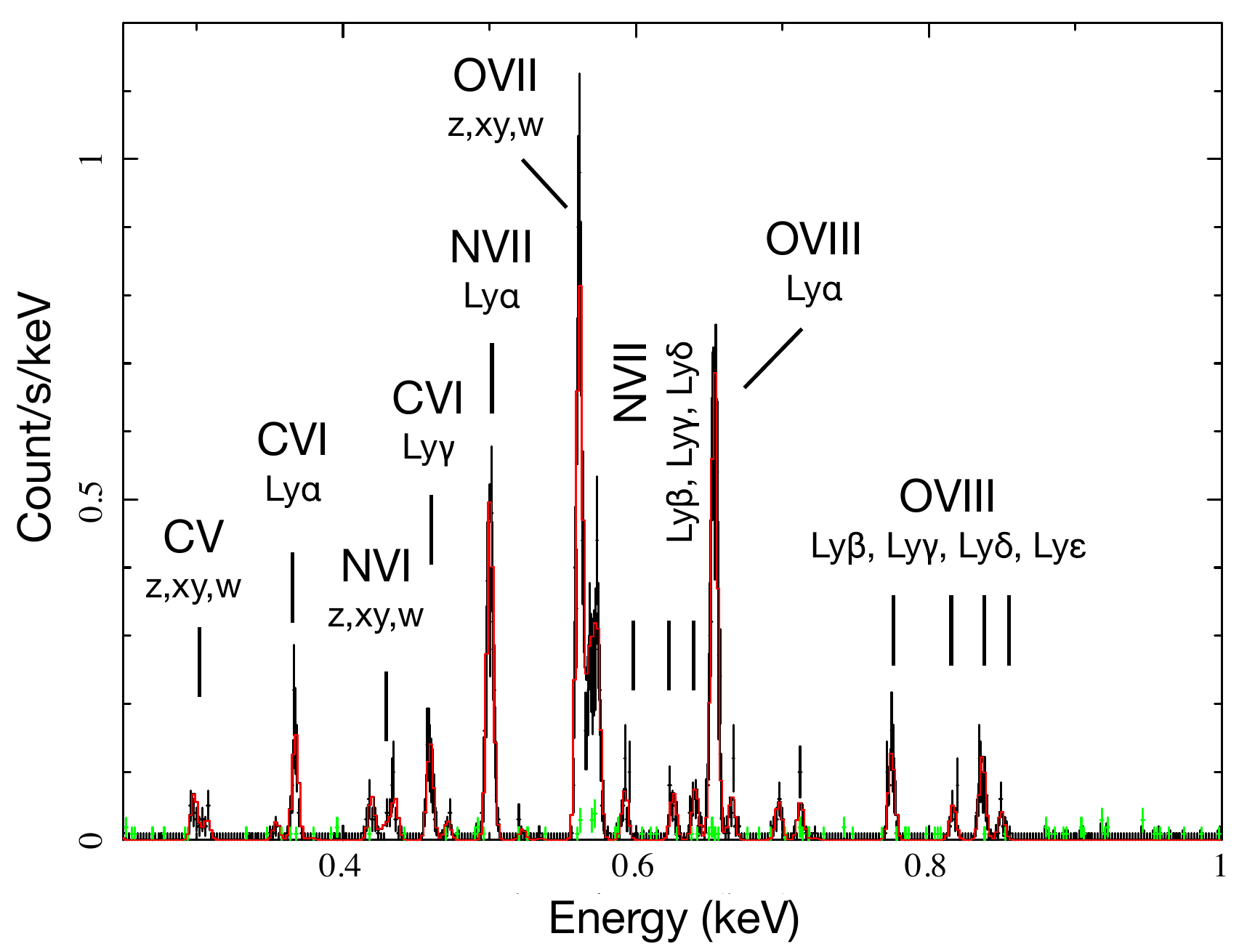}
\caption{A 50~ks SXS simulation of the 
strong
geocoronal SWCX event. Green shows a simulated 50 ks spectrum of SXB for comparison.}
\label{fig:geocorona-simulation}
\end{floatingfigure}
Because the geocoronal SWCX always exists as background emission in all {\it ASTRO-H} observations, 
a dedicated observation time is not necessarily needed. By carefully checking short time variability 
on the order of hr or day of emission lines and comparing the X-ray data with the solar wind data, we can pick up the SWCX event.
By searching 6-yr {\it Suzaku} archive data for time variable geocoronal SWCX events, \cite{ish13} found 
that about $\sim$2 \% data are affected by the geocoronal SWCX. The OVII line intensity, 
which is one of the most prominent lines in the SWCX, averaged over each observation, ranges from $\sim$3 
to $\sim$50 Line Units (LU),  where a LU is defined as a photon s$^{-1}$cm$^{-2}$sr$^{-1}$.
From this study, we expect about 5 strong geocoronal SWCX events per year with the OVII line 
intensity more than $>$10 LU as long as the solar activity after launch (2015$\sim$16) 
is as high as that in 2005. This is highly likely considering the 11 yr cycle of the solar activity.

As an example, a 50 ks observation of the 
strong
geocoronal SWCX event is simulated based on 
\cite{ezo11b}, in which the OVII line intensity was 34 LU, as shown in Figure 
\ref{fig:geocorona-simulation}.
The exposure time corresponds to a typical duration of a large magnetic storm during which the solar wind flux significantly increases. 
The assumed SWCX model originally constructed by \cite{bod07} accounts for the relative emission 
cross sections for the lines from each of several ions. Emission lines from CV, CVI, 
NVI, NVII, OVII, and OVII are clearly detected.
Although the solar wind ion abundance can change event by event, for strong emission line(s), 
the line center energy, width, shift and intensity can be determined with high accuracy of 
the order of $\sim$$\pm$1 eV and $\sim$10\%.

For comparison, we simulated a 50 ks spectrum of the soft X-ray background (SXB) 
consisting the CXB component represented by an absorbed broken power-law function,
and two thin-thermal emission components to represent the heliospheric SWCX induced
emission lines plus Local Hot Bubble and Galactic emission from \citet{yos09}. 
We can confirm that photons from the geocoronal SWCX dominate the data. 

\noindent
{\bf Beyond Feasibility:}

Similar to the planetary CX emission, a TOO joint observation with solar wind monitoring 
satellites maybe possible to aim at detection of a extremely bright SWCX event. A blank sky 
region such as the north ecliptic pole will be proper as a line of sight direction, to avoid 
contamination from bright soft X-ray sources and to study the magneto-cusp region.


\subsection{Comets}

\noindent
{\bf Background and Previous Studies:}

\begin{figure}
\begin{center}
\includegraphics[width=0.8\hsize]{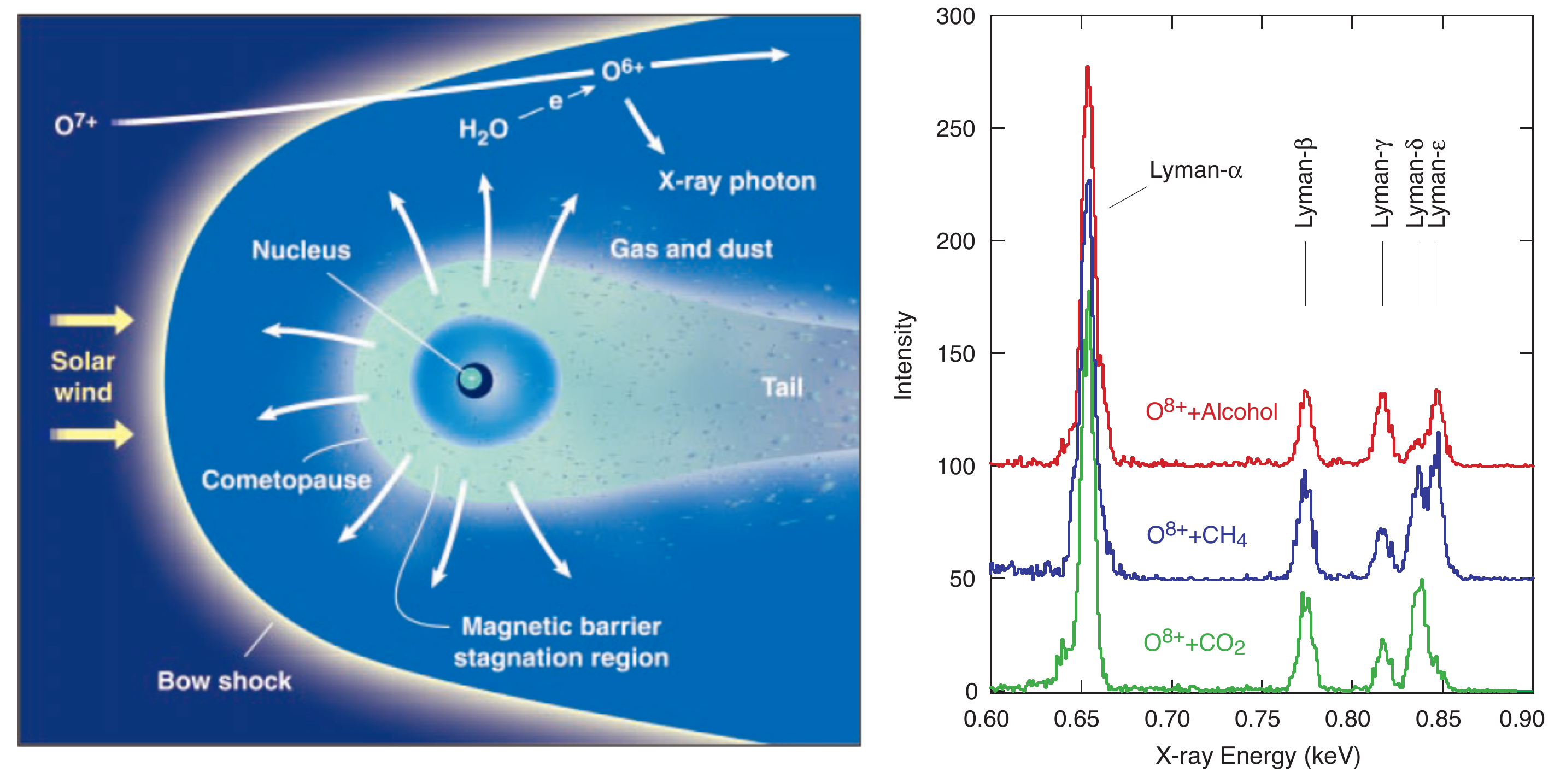}
\end{center}
\vspace*{-5mm}
\caption{
[Left] Schematic of cometary SWCX \citep{cra97}.  [Right] Laboratory experiment showing K-shell emission 
from O$^{7+}$ after charge exchange with different molecules.
\citep{bei03}.
}
\label{fig:comet-concept}
\end{figure}

The discovery of soft X-ray emission from Comet Hyakutake has revealed the existence of a new class 
of X-ray emitting objects, i.e., SWCX between the solar wind and neutrals \citep{lis96}.
A widely spread coma, whose size is 10$^{4-6}$~km and the number density ranges 10$^{0-8}$~molecules~cm$^{-3}$ 
depending on the distance from the core of the coma (e.g., \citealt{lis03}), emits X-rays via SWCX, as shown
in Figure \ref{fig:comet-concept}[left].
Since the comets are one of the brightest X-ray sources in the solar system, observational studies of the cometary 
X-ray emission must enable us to a better understanding of not only the cometary coma but also the SWCX process 
itself which is known to play an important role in other astrophysical objects. 

\smallskip
\noindent
{\bf Prospects \& Strategy:}

Similar to the geocoronal and heliospheric SWCX, for the first time with the {\it ASTRO-H} SXS, the high energy 
resolution spectroscopy will be possible for comets, since they are diffuse X-ray objects. 
Line properties such as Doppler shift, broadening if exists, and intensity will provide us with information
on the dynamics of ion, the solar wind ion abundance, and the neutral density in the cometary coma, in the
same way as the other SWCX sources. 

A ground-based experiment using the ASTRO-E XRS spare detector revealed the potential for using cometary X-rays to infer information about the chemical composition of the cometary gases \citep{bei03}. The principal quantum number of electron capture changes with the ionization potential of the electron-donating material.  Electrons will be captured into higher n-levels if the ionization potential of the donor electron is lower, resulting in different high n-level line distributions. The experimental result is shown in Figure 
\ref{fig:comet-concept}[right].

\smallskip
\noindent
{\bf Beyond Feasibility:}

Similar to the other SWCX sources, a huge increase of the X-ray flux can be expected according 
to an increase of the incident solar wind associated with such as coronal mass ejection and 
co-rotational interaction region. 

\subsection{The ``Beyond''}

X-rays from CX may also be observable from more distant sources such as 
supernova remnants and galaxies.  CX emission in SNRs are thought to play 
an enhanced role where neutrals are mixed with shocked hot gas.  In fact, 
observational evidence of CX emission was found at optical wavelengths 
more than 30\,yrs ago \citep[e.g.,][]{Kirshner1978}.  In the X-ray domain, 
\citet{Wise1989} first performed detailed calculations of CX-induced X-ray 
emission in a SNR.  They found that CX emission generally contributes only 
10$^{-3}$ to 10$^{-5}$ of the collisionally excited lines in the entire 
SNR.  On the other hand, \citet{Lallement2004} later examined projected 
emission profiles for 
both CX and thermal emission in SNRs, and noted that CX X-ray 
emission may be comparable with thermal emission in thin layers at the 
SNR edge.  It is also pointed out that the relative importance of CX X-ray 
emission to thermal emission is proportional to a quantity, 
$n_\mathrm c$~$V_\mathrm r$~$n_\mathrm e^{-2}$, with $n_\mathrm c$ being 
the cloud density, with $V_\mathrm r$ the relative velocity between 
neutrals and ions, and $n_\mathrm e$ the electron density of the 
hot plasma.  Thus, the higher the density contrast 
($n_\mathrm c$/$n_\mathrm e$), the stronger the presence of CX X-ray 
emission would become.

There are growing observational hints of CX X-ray emission in SNRs, 
1E0102.2--7219 in the SMC \citep{Rasmussen2001}, the Cygnus Loop 
\citep{Katsuda2011} and Puppis~A \citep{Katsuda2012} in our Galaxy.
The line intensity ratios of 
($n>2 \rightarrow n=1$)/($n=2 \rightarrow n=1$) in H-like O 
(1E0102.2--7219) or He-like O (Cygnus Loop) seem to be higher than 
expectations from thermal emission models.  In Puppis~A, 
forbidden-to-resonance line ratios in He$\alpha$ transitions derived 
by the {\it XM-Newton} RGS spectra are found to be anomalously enhanced 
at the most prominent cloud-shock interaction regions.  These could be 
naturally explained by the presence of CX emission, although other 
possibilities such as uncertainties on Fe L-shell lines, missing 
inner-shell satellite lines, and effects of resonance line scattering 
as well as recombination.

In star-forming regions, hot plasmas generated by massive stellar 
clusters can meet cold neutral gases, as they move through crevices of 
cold clouds.  At the interfaces, highly-ionized ions in the hot plasma 
can interact with neutrals, and thus CX X-rays can be emitted.  In fact, 
during the recent {\it Chandra} Carina Complex Project, it is noticed 
that the X-ray spectrum from a southern outflow in a giant HII region, 
NGC~3576, requires a Gaussian at 0.72\,keV to obtain a good fit.  This 
Gaussian component has been interepreted as CX X-ray emission 
\citep{Townsley2011}, although other possibilities are not yet ruled out,
e.g., Fe L lines missing in the current plasma codes or something else 
due to previously-unrecognized physical processes.

By analogy to the star-forming regions, star-forming galaxies are also 
possible sites where CX X-ray emission can play an important role.  
A marginal detection of an emission line at 0.459\,keV is reported for 
the starburst galaxy M82, which may be due to the CX emission of C VI 
\citep{Tsuru2007}.  Stronger supports for the CX were later found in the 
line ratios of the O VII He$\alpha$ \citep{Ranalli2008,Liu2011}.
Recent high-resolution X-ray spectroscopy with the {\it XMM-Newton} RGS has 
been revealing fine structures of O VII He$\alpha$ 
lines for a number of bright central regions of star-forming galaxies.  
For most cases, the forbidden lines are found to be comparable to or 
even stronger than the resonance lines \citep{Liu2012}.  Together with 
the correlation with H$\alpha$ emission and the X-ray emission 
\citep[e.g.,][]{Wang2001}, this is best explained by significant 
contributions of CX emission.  The CX X-ray emission may be 
ubiquitous for galaxies with massive star formation.

The non-dispersive spectrometer SXS on board {\it ASTRO-H} will allow 
for further tests for the CX scenario and other possibilities.  
The SXS is able to reveal spatial distributions of line intensity ratios 
such as forbidden-to-resonance in K$\alpha$ transitions and 
($n>2 \rightarrow n=1$)/($n=2 \rightarrow n=1$) that are key information 
to find CX signatures in thermally-dominated X-ray emission.  
The CX scenario can then be examined by seeing if there 
are any spatial correlations between these lines ratios and the distance 
from the shock front and/or locations of optical clouds which could be 
electron donors.  The SXS is also sensitive to other complex and/or 
faint spectral features such as radiative recombination continua, when 
compared with the RGS having smaller effective area and suffering from energy 
degradation due to the spatial extent of the source.  
In this way, the SXS will be essential to examine the possibilies of  the CX X-ray 
emission in many astrophysical sites including SNRs, star-forming regions, and 
galaxies.  Taking account of the CX emission would be extremely important for 
understanding the true properties of the shocked plasma, since the undetected 
presence of CX emission leads to incorrect plasma parameters (e.g., the 
metal abundance and the electron temperature) if we interpret the 
CX-contaminated X-ray spectra with pure thermal emission models.  
An SXS simulation for the Cygnus Loop SNR can be found in the {\it ASTRO-H} 
white paper (Knox et al. 2013).

\section{Advanced Spectral Modeling: Photoionized and Collisional Plasmas
\label{sec:nsf/advanced_modeling}}


\subsection{Background and Previous Studies}
\label{advance:bgd}

X-ray astronomy, particularly for bright sources, will change dramatically once high-resolution spectra from {\it ASTRO-H} become available. 
We will encounter a number of unexpected spectral features with unprecedented statistics. In 
such cases, the simple method of `chi-squared fitting' (relying on 
canned spectral models with a small number of parameters) will no longer return reliable 
results, and requiring an advanced modeling effort.

\begin{figure}
  \begin{center}
      \includegraphics[width=0.8\hsize]{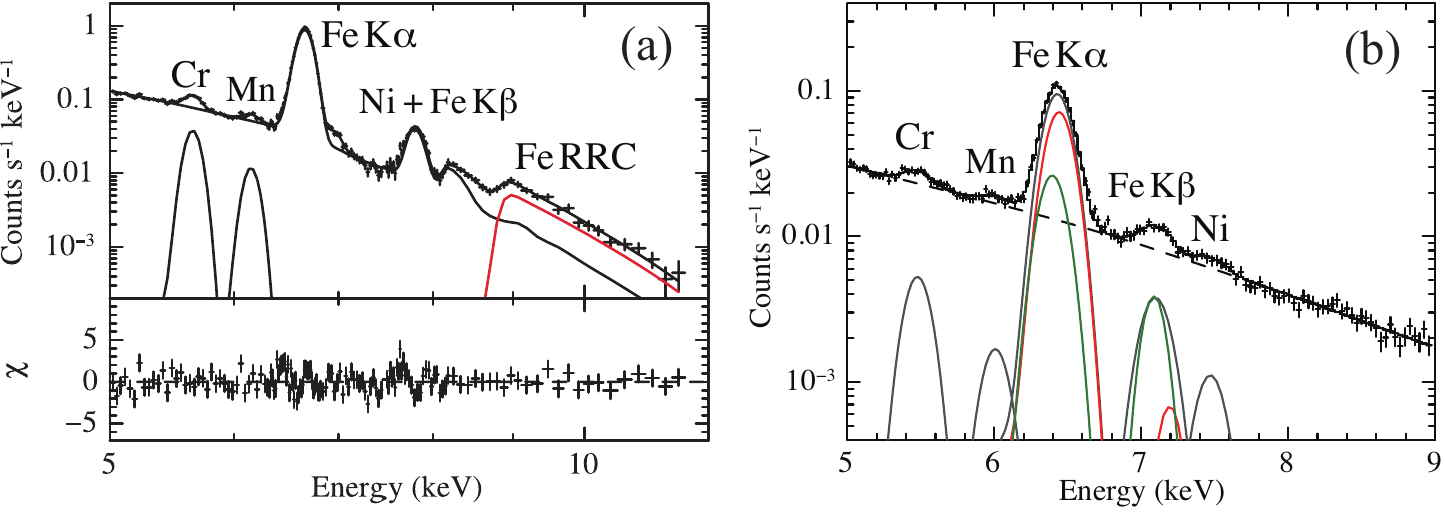}
  \caption{(a) {\it Suzaku} spectrum of W49B (Ozawa et al.\ 2009). 
  The red line represents the strong RRC of He-like Fe, which can be enhanced only when the plasma is overionized. \
    (b) {\it Suzaku} spectrum of the NW rim of Tycho \citep{Yamaguchi13}. 
    The K$\alpha$ and K$\beta$ emission lines of Fe are successfully reproduced by 
    a `two-component' model consisting of low-ionization (around Fe$^{8+}$: green) and 
    higher-ionization (around Fe$^{16+}$: red) components.  The K$\beta$/K$\alpha$ 
    ratios of the former and latter are assumed to be 0.15 and 0.01, respectively. 
    The low ratio for Fe$^{16+}$ is expected because there is no bound electron 
    in the $3p$ shell. The detailed values of the centroids and intensity ratio 
    for each charge stage is given in Table\,\ref{tab:fe}.}
  \label{suzaku}
  \end{center}
\end{figure}

Unexpected spectral results have already been obtained even with recent moderate-resolution 
data with {\it Suzaku}, taking advantage of deep observations with low background. The 
discovery of strong radiative recombination continua from collisionally-ionized plasmas 
\citep{Yamaguchi09, Ozawa09} as shown in Figure\,\ref{suzaku}a was one of the most 
surprising results concerning supernova remnants (SNRs), has and significantly changed our 
understanding of SNR's dynamics and evolution (see also the Old SNR WP). 
{\it Suzaku}'s low-background and good resolution has also enabled us to detect low-
abundance elements 
(e.g., Al, Cr, Mn, Ni) and weak emission lines from more abundant elements (e.g., Fe K$\beta$), providing 
useful diagnostics of the progenitor's nature \citep[e.g.,][]{Badenes08} and ion population in 
plasmas (Yamaguchi et al.\ 2013, see Figure\,\ref{suzaku}b and its caption). It is notable that 
most of these spectral features were not predicted by any public model at the time of the 
discoveries.  This is a key point: Do not rely too much on prepared spectral models someone 
else made! There is no model free of assumptions, limitations, or applicable ranges.  The high-
resolution data of {\it ASTRO-H} will, more than often, show unexpected and important features, 
and a shift to an advanced modeling approach based on a knowledge of fundamental atomic 
physics will be needed. Otherwise, we may overlook important signatures that the 
observational data will be telling us. For example, although new SWCX models are now 
available, they are only applicable in specific circumstances; applying them without care will 
lead to unphysical results that cannot be justified even if they fit parts of a spectrum.

\subsection{Non-Equilibrium Ionization and Ion Population}
\label{advance:nei}

\begin{figure}
\centering \includegraphics[width=0.6\hsize]{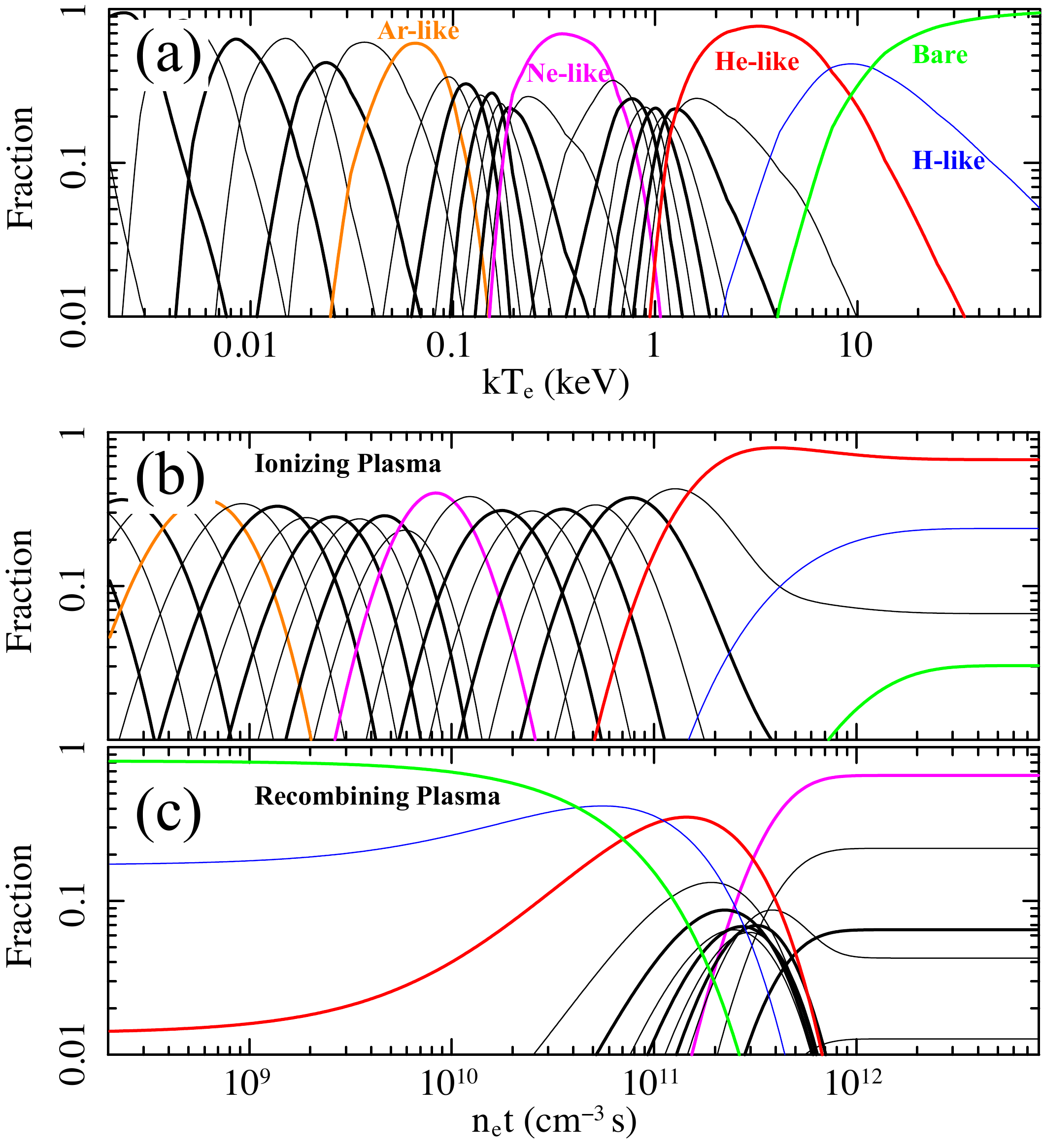}
  \caption{(a) Ion population of Fe in a CIE plasma as a function of the electron temperature. 
  Orange, magenta, red, blue, and green represent Fe$^{8+}$ (Ar-like), 
  Fe$^{16+}$ (Ne-like), Fe$^{24+}$ (He-like), Fe$^{25+}$ (H-like), 
  and Fe$^{26+}$ (fully-stripped), respectively. \ (b) Ion population of Fe 
  in an $kT_e = 5$\,keV ionizing plasma as a function of the ionization age. \ 
  (c) Same as (b), but for recombining plasma where the initial and current 
  electron temperatures are assumed to be 30\,keV and 0.3\,keV, respectively.}
  \label{ionpop}
\end{figure}

In this and the next subsections, we discuss collisionally-ionized plasmas where both excitation 
and ionization by UV/X-ray photons are negligible.  At an X-ray-emitting temperature, heavy 
elements in plasmas are highly ionized, usually to He-like or H-like state (except for the 
heaviest abundant elements such as Fe and Ni). This is because the internal kinetic energy of 
electrons are comparable to the K-shell ionization potential of the astronomically-abundant 
heavy elements (i.e., C through Ni). Figure\,\ref{ionpop}a shows the ion population of Fe as a 
function of electron temperature $kT_e$, assuming ionization equilibrium (CIE). In a CIE 
plasma, the fractional abundance of each ion can be determined uniquely by $kT_e$, so 
spectral analysis is relatively straightforward --- just applying a simple CIE plasma model (e.g., 
{\it apec}: \citet{Smith01, Foster12}; {\it mekal}: \citet{Mewe85, Kaastra92}) would be fine (but 
we should still be careful about uncertainty and limitation in the atomic data). The assumption 
of CIE is fulfilled in a majority of astronomical X-ray emitting plasmas, such as galaxy clusters. 

\begin{figure}
  \begin{center}
      \includegraphics[width=0.6\hsize]{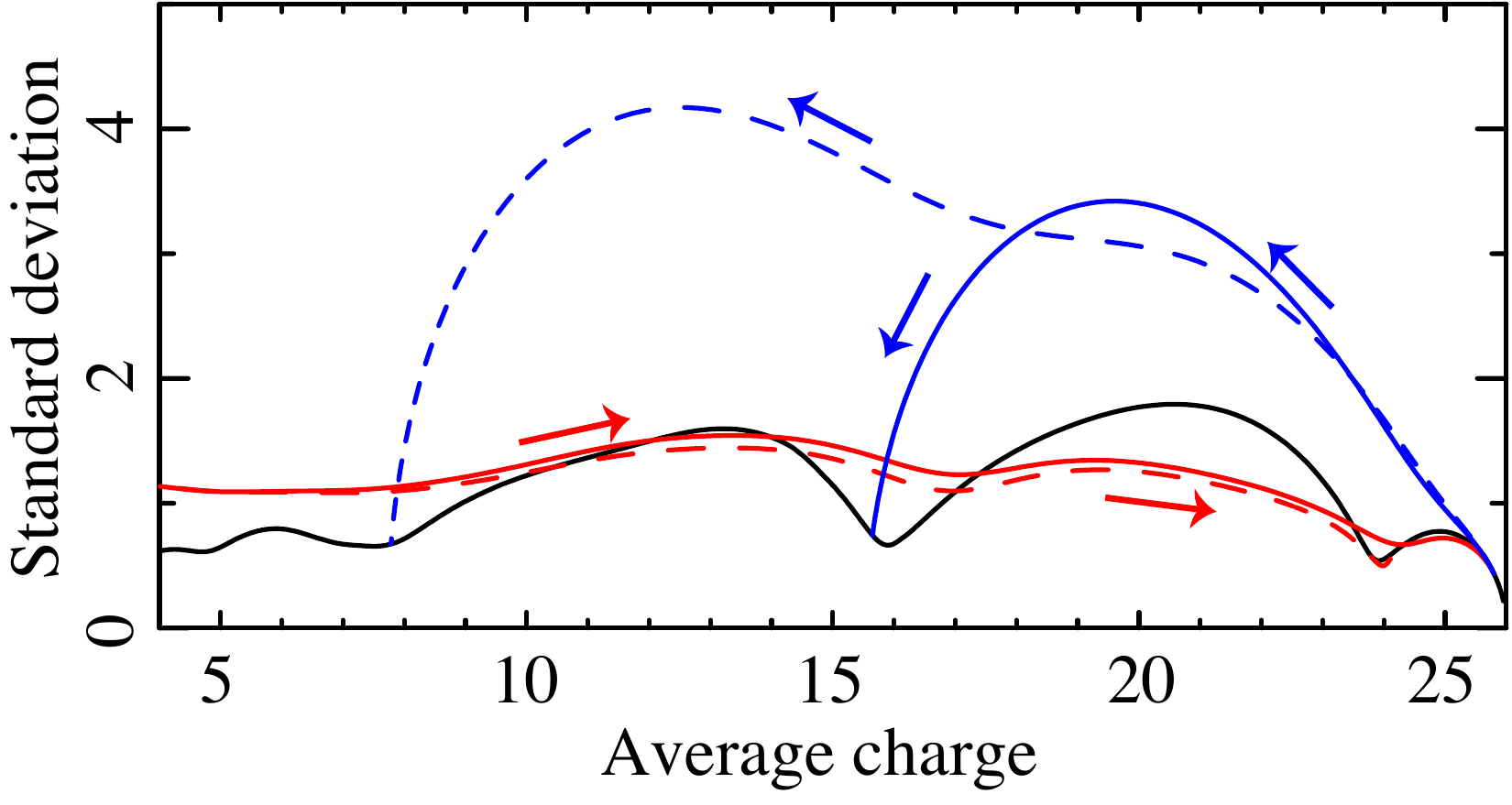}
  \caption{Standard deviation of a charge number vs average charge for 
  a CIE (black), ionizing (red), and recombining (blue) plasmas. 
  }
  \label{deviation}
  \end{center}
\end{figure}

However, the situation is more complicated when the physical condition (e.g., temperature) of 
the plasma suddenly changes and the timescale of the temperature fluctuation is shorter than 
that required to reach CIE. Such an unsteady condition is called ``Non-Equilibrium Ionization 
(NEI)''. An NEI plasma is commonly observed in SNRs as they form shock waves heating ISM 
or supernova ejecta up to high temperature abruptly. The degree of NEI is quantified by 
``ionization age'' parameter $\tau = n_e t$, a product of electron 
density ($n_e$) and elapsed time since the sudden temperature change ($t$).  The timescale 
to reach CIE depends on the electron temperature and element, and is typically in a range of 
$n_et = 10^{10} - 5 \times 10^{12}$\,cm$^{-3}$\,s \citep[{\it e.g.,}][]{SmithHughes10}. 
Figure\,\ref{ionpop}b gives ion fraction in a $kT_e = 5$\,keV ionizing NEI plasma as a 
function of $n_et$. The ion population calculated for an ionizing plasma 
largely differs from that in a CIE plasma. As expected, the charge number of the ions is very 
low (almost neutral) in the early NEI stage and gradually goes up as the $n_et$ value 
increases. It is worth noting that, in the CIE plasma, the abundance curve of the Fe$^{16+}$ 
(Ne-like) ion shows a `plateau' (relatively flat region) at $\sim$0.3--0.6\,keV, while those of the 
non-closed-shell ions (e.g., Fe$^{12+}$, Fe$^{20+}$) have a steeper peak.  In the ionizing 
plasma, on the other hand, the shape of the Fe$^{16+}$\ abundance curves looks very similar 
to those of the other ions. This can be more quantitatively presented in Figure\,\ref{deviation}. 
This shows the standard deviation of a charge number 
$\sigma = [\Sigma_{i=0}^{26} \, f_i \, (z_i \, - \,$$<$$z$$>$$)^2]^{1/2}$ in both cases, 
where $f_i$, $z_i$\ and $<$$z$$>$\ are the fraction and charge number of Fe ion (Fe$^{z{_i}+}$) 
and the average charge, respectively. Unlike the CIE case, relatively flat deviations are 
achieved throughout the evolution of the ionizing plasma. 

The ion populations expected for a recombining plasma are strikingly different from those in the 
CIE and ionizing plasmas. Figure\,\ref{ionpop}c demonstrates evolution of a $kT_e = 0.3$\,keV 
recombining plasma assuming that the initial ionization balance corresponds to a 30\,keV 
CIE plasma. Interestingly, the fully-stripped and Ne-like ions can co-exist during a certain 
evolutional stage (which cannot happen either in the CIE or ionizing case), and also the 
standard deviation is expected to be much larger than those in the other cases (Figure\,
\ref{deviation}). Such a broad distribution of different ions can be achieved because 
the recombination rate does 
not strongly depend on the charge number, in contrast to the ionization rate.  

The difference in the expected ion population among the different plasma states (i.e., CIE, ionizing, recombining) becomes rather important when we introduce a physical quantity, ``ionization temperature ($kT_z$)'' which can be defined for a particular ionization distribution as the temperature such that the {\it average charge} of the ionization distribution is consistent with that of a CIE plasma with electron temperature $kT_z$.  
We should note, however, that although the average charge of the {\it ion population} of a $kT_z = 1$\,keV 
recombining plasma is by definition identical to that of a $kT_e = 1$\,keV CIE 
plasma, the overall ion distribution will be hugely different\footnote{Therefore, an assumption of 
a CIE-like ionization balance, introduced in several previous works, is not appropriate for 
analysis of a NEI plasmas unless they are quite close to being in equilibrium. This approach 
can lead inaccurate spectral profiles and elemental abundances.}. Moreover, in realistic 
ionizing or recombining plasmas, an electron temperature also 
evolves with time and both $kT_e$ and $n_et$ have spatial non-uniformity. Therefore, ion 
population in the realistic sources will be not as simple as the simple NEI models predict. In 
fact, it was found that the {\it Suzaku} spectrum of Tycho (Figure\,\ref{suzaku}b) cannot be 
represented by a single-component NEI or parallel-shock model \citep[; see the Young SNR 
WP for more detail][]{Yamaguchi13}.  Future spectral analysis especially for high-resolution 
data should, therefore, be more flexible so that an arbitrary ion population can be described.

\subsection{Innershell Process and Fluorescent X-Ray Emission}
\label{advance:inner}

\begin{figure}
  \begin{center}
      \includegraphics[width=0.9\hsize]{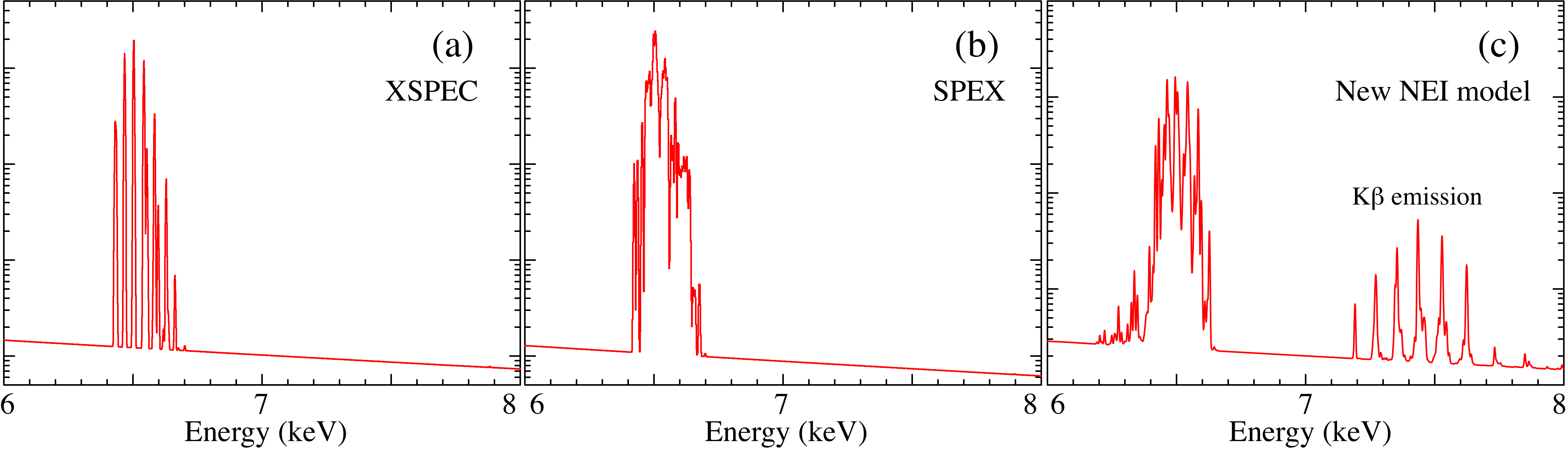}
  \caption{Comparison of model spectra for an ionizing plasma with $n_et = 2 \times 10^{10}$\,cm$^{-3}$\,s, predicted by XSPEC, SPEX, and detailed calculation using the ``Flexible Atomic Code'' \citep{Eriksen13}. 
  No K$\beta$ emission is involved in XSPEC or SPEX despite its importance for diagnostics.}
  \label{compare}
  \end{center}
\end{figure}

Innershell ionization/excitation processes are particularly important for high-$Z$ elements in young SNRs, since their ionization stage is usually observed to be lower than the He-like state. Nonetheless, there is currently no public model that completely includes these processes due to a inadequate atomic data. Figure\,\ref{compare} compares model spectra of an NEI plasma with $n_et$ of $2 \times 10^{10}$\,cm$^{-3}$\,s predicted by XSPEC, SPEX, and a more accurate calculation using the ``Flexible Atomic Code (FAC)'' \citep{Eriksen13}.  The most obvious difference is seen in the energy range between 7 and 8\,keV.  We find neither the XSPEC nor SPEX NEI models include any K$\beta$ emission except for He-like and H-like states. The energy (wavelength) of K$\alpha$ emissions in the existing models is also found to be inaccurate; for example, all the K$\alpha$ emissions of Fe\,{\footnotesize{\rm II--XII}} and Fe\,{\footnotesize{\rm XIII--XVII}}\footnote{We caution that, in a number of papers, expressions such as ``Fe\,{\tiny{\rm XVII}}'' and ``Fe$^{16+}$'' are used interchangeably.  The correct usage is straightforward: epressions with a roman number indicate {\it emission} (or {\it absorption}\ bound-bound transition lines, while that with a superscript number indicates {\it ions}.  Therefore, the emission expected after the innershell ionization of Fe$^{16+}$ is Fe\,{\tiny{\rm XVIII}}, not Fe\,{\tiny{\rm XVII}}. Also, an expression like O\,{\tiny{\rm IX}} is totally incorrect, because there can be no line emission from fully-stripped oxygen.} are  placed at 6390\,eV and 6425\,eV, respectively, in SPEX (see Table\,\ref{tab:fe} for comparison with the accurate values), while the XSPEC NEI model obviously oversimplifies the line profiles as shown in Figure\,\ref{compare}a. In order to determine accurate ionization ages, abundances, and ion temperatures, accurate predictions of the line energy and emissivity is essential. Table\,\ref{tab:fe} gives the centroid energies of the Fe K$\alpha$\ and K$\beta$\ emissions and their intensity ratio for different charge numbers at $kT_e = 5$\,keV. We emphasize that the modeling of K$\beta$ emission is particularly important because: (1) the centroid energies of the Fe K$\beta$ lines are well separated from those of the next charge stages, compared to the Fe K$\alpha$ lines, which will be an advantage when measuring ion fractions and thermal-Doppler line widths (see Figure\,\ref{compare}c). (2) the K$\beta$/K$\alpha$ intensity ratio is sensitive to the charge state, and thus a diagnostic of the ion population \citep[][; see also the Young SNR WP]{Yamaguchi13}. Since the K$\beta$/K$\alpha$ ratio becomes higher than 10\% at $z \leq 10$, clear detection of K$\beta$ emission is expected from most SNRs with low-ionized Fe ejecta (e.g., Tycho, Kepler, SN1006, Type Ia SNRs in the LMC). For highly-ionized Fe ($14 \leq z \leq 18$), on the other hand, the ratio goes down to $\lesssim 2$\%. However, their contribution is still the subject of consideration, because their line energies overlap with those of Ni-K$\alpha$ emission (which is also weak compared to Fe-K$alpha$ lines). The accurate Fe-K$\beta$ data will, therefore, help accurate measurement of Ni abundances.

\begin{table}
\begin{center}
\begin{tabular}{lcccclccc}
\hline \hline
$z$ &$E_{{\rm K}\alpha}$ (eV) & $E_{{\rm K}\beta}$ (eV) & K$\beta$/K$\alpha$ ratio & \ \ \ \ \ \ 
& $z$ & $E_{{\rm K}\alpha}$ (eV) & $E_{{\rm K}\beta}$ (eV) & K$\beta$/K$\alpha$ ratio \\
\hline
0 & 6402 & 7059 & 0.120  && 12 & 6415 & 7144 & 0.063 \\ 
1 & 6402 & 7060 & 0.121  && 13 & 6420 & 7158 & 0.036 \\ 
2 & 6402 & 7060 & 0.122  && 14 & 6420 & 7153 & 0.015 \\
3 & 6401 & 7059 & 0.127  && 15 & 6425 & 7176 & 0.010 \\ 
4 & 6400 & 7063 & 0.132  && 16 & 6427 & 7192 & 0.012 \\ 
5 & 6399 & 7070 & 0.136  && 17 & 6455 & 7270 & 0.009 \\
6 & 6399 & 7075 & 0.141  && 18 & 6484 & 7351 & 0.013 \\ 
7 & 6399 & 7081 & 0.149 && 19 & 6517 & 7434 & 0.025 \\ 
8 & 6398 & 7090 & 0.168 && 20 & 6544 & 7517 & 0.029  \\ 
9 & 6401 & 7102 & 0.146 && 21 & 6575 & 7610 & 0.036  \\ 
10 & 6405 & 7116 & 0.122 && 22 & 6589 & 7705 & 0.044  \\ 
11 & 6410 & 7128 & 0.099 && 23 & 6641 & 7777 & 0.075  \\ 
\hline
\end{tabular}
\caption{The centroid energies of Fe-K$\alpha$ and K$\beta$ 
emissions and their intensity ratios for the different charge numbers $z$. 
For $z = 9-13$, only K-shell ionization (and no excitation) is taken into 
account, so the centroids are slightly overestimated and 
the ratios are underestimated. More detailed information/values will be 
found in \citet{Yamaguchi13}.} 
	\label{tab:fe}
\end{center}
\end{table}

Innershell ionization of Li-like ions frequently creates the excited $1s2s$ state, and hence can significantly contribute to the formation of the He-like forbidden line. Therefore, not only the $G$ ratio ($\equiv (x+y+z)/w$) but also the $R$ ratio ($\equiv z/(x+y)$) is expected to be enhanced \citep[e.g.,][]{MeweSchrijver85}. This provides a useful diagnostic to distinguish the innershell effect from other processes that can also induce the forbidden transition ({\it i.e.}, radiative/dielectronic recombination and charge exchange).  Both the recombination and charge exchange processes favor the populations of the triplet levels, due to their higher statistical weight compared to the singlet level. Therefore, the intercombination lines will also be enhanced, resulting in a smaller $R$ ratio.

Most (probably all) of the public NEI codes use values of fluorescence yields ($\omega \equiv A_r/(A_r + A_a)$, where $A_r$ and $A_a$ are radiative and Auger rates, respectively) calculated by \citet{KaastraMewe93}. This work used a single-configuration approach in order to keep the calculation to manageable size for the time and thus several important effects, such as configuration interaction, were ignored.  It has been pointed out that this simplification can be problematic in some situations \citep[e.g.][]{Gorczyca06}. One of the most significant cases is seen in the Li-like $1s2s^2$ state (after innershell ionization of a Be-like ion). Since this excited state contains no $2p$ electron, a permitted (electric dipole) transition is impossible. Therefore,\citet{KaastraMewe93} gave $\omega = 0$\ for every Li-like ion. However, once the configuration interaction effect is taken into account, the yields will have non-zero values due to 
mixing with a $1s2p^2$ wave function. This effect becomes prominent especially for high-$Z$ elements. 
For example, the yield was found to be $\sim$0.12 for $1s2s^2$ Fe$^{23+}$ \citep{Gorczyca06}, so the fluorescent emission significantly contributes to a spectrum. As an example, this amounts to $\sim 10\%$\ increase in the Li-like line flux for $n_e t = 7\times10^{10}$~cm$^{-3}$~s where Be-like Fe is dominant.  We should, therefore, properly take into account these effects in order to obtain an accurate prediction of the satellite line profile. 

In addition to the inner K-shell ionization and fluorescence, L-shell fluorescence could also be important for low-ionized plasmas. The current plasma models are designed to have Fe L-shell emission only above Fe$^{16+}$. This is because at typical X-ray-emitting temperatures ($> 0.2$\,keV), the population of Fe$^{15+}$ 
and lower-charged Fe is almost negligible (Figure\,\ref{ionpop}a). However, this is not the case for an ionizing plasma with an extremely low ionization age ($n_et < 5 \times 10^9$~cm$^{-3}$\,s; see Figure\,\ref{ionpop}b), and thus there is no reason that we ignore the L-shell emissions from the lowly-charged Fe. 
A comprehensive model that includes these missing lines should be established for high-resolution spectroscopy with {\it ASTRO-H}. 

\subsection{Photoionized Plasma}
\label{subsec:nsf/advanced_modeling/photoionized_plasma}

As well as the collisional ionization discussed above, photoionization is an important process that drives X-ray emitting plasmas. Photoionized plasmas driven by X-rays, which are commonly seen around accretion-powered objects such as neutron stars and black holes including active galactic nuclei (AGN), reveal the physical conditions of the surroundings of the energetic objects through complicated spectra containing a number of emission or absorption lines.  This information will play a key role in revealing how the accretion power affects the surrounding environment, leading to the knowledge of the evolution of galaxies ({\it See AGN chapter}).  However, modeling of the photoionized plasma is highly complex since the physical conditions of the plasma cannot be determined purely by local properties. As the plasma is driven by an external X-ray source,  the geometrical relation to the source and even to other parts of the plasma must be considered.

To calculate the spectrum of a photoionized plasma, there are three steps to obtain the results:
\begin{enumerate}
\item understand physical processes from a microscopic viewpoint;
\item determine physical states of the plasmas; and
\item calculate radiative transfer. 
\end{enumerate}
For the first item, we need models of emissions and photon interactions with ions based on atomic physics and precise atomic data.
The basic points of this problem should be common to those for the thermal plasmas discussed in the previous subsections.
Nevertheless, the difference in the efficiency of atomic processes is worthy of attention.
For example, in photoionized plasmas, dielectronic recombination is negligible in many cases since radiative recombination is much more efficient at low temperature than dielectronic recombination.
Such a disregard simplifies the calculation of the ionization balance in the next step.
The second step is to determine hydrodynamical structure of the plasma including temperatures, densities, velocities, and ion abundances (chemical compositions and ion fractions).
Although these properties can be all coupled, it is usually sufficient to calculate the distributions of ion fractions and temperatures by balancing ionization and recombination, and also balancing heating and cooling with the assumption that other hydrodynamical properties such as the density distribution can be fixed. The third step gives the resultant emission from the plasma. Since the radiative transfer also affects the photoionization and photoexcitation processes, the second and third steps are strongly coupled, and the problem should be solved self-consistently. Many photoionization calculating codes, e.g.\ XSTAR \citep{Kallman:1982}, CLOUDY \citep{Ferland:1998}, have adopted iterative algorithms to obtain the final solution.

As the problem of radiative transfer is a major issue in astrophysics, we discuss it separately in the next section, and here we briefly describe a strategy.  If the system is optically thin, the degree of photoionization is easily characterized by a ionization parameter $\xi$ defined as
\begin{equation}
\xi=\frac{L_X}{n_e r^2}=\frac{4\pi F_X}{n_e},
\end{equation}
where $L_X$, $n_e$, and $r$ denote the X-ray luminosity of the source, the electron density, and the distance to the source, respectively. This parameter controls the ionization balance since the ionization and recombination rates are proportional to the ionizing flux $F_X$ and the electron density $n_e$, respectively. When the system becomes optically thick, one-dimensional radiative transfer in a simplified geometry such as plane-parallel geometry or spherical geometry provides an effective approach. In more complicated case, a Monte Carlo simulation would be preferred \citep{Watanabe:2006, Sim:2008}.

\begin{figure}[htbp]
\begin{center}
\includegraphics[width=0.95\hsize]{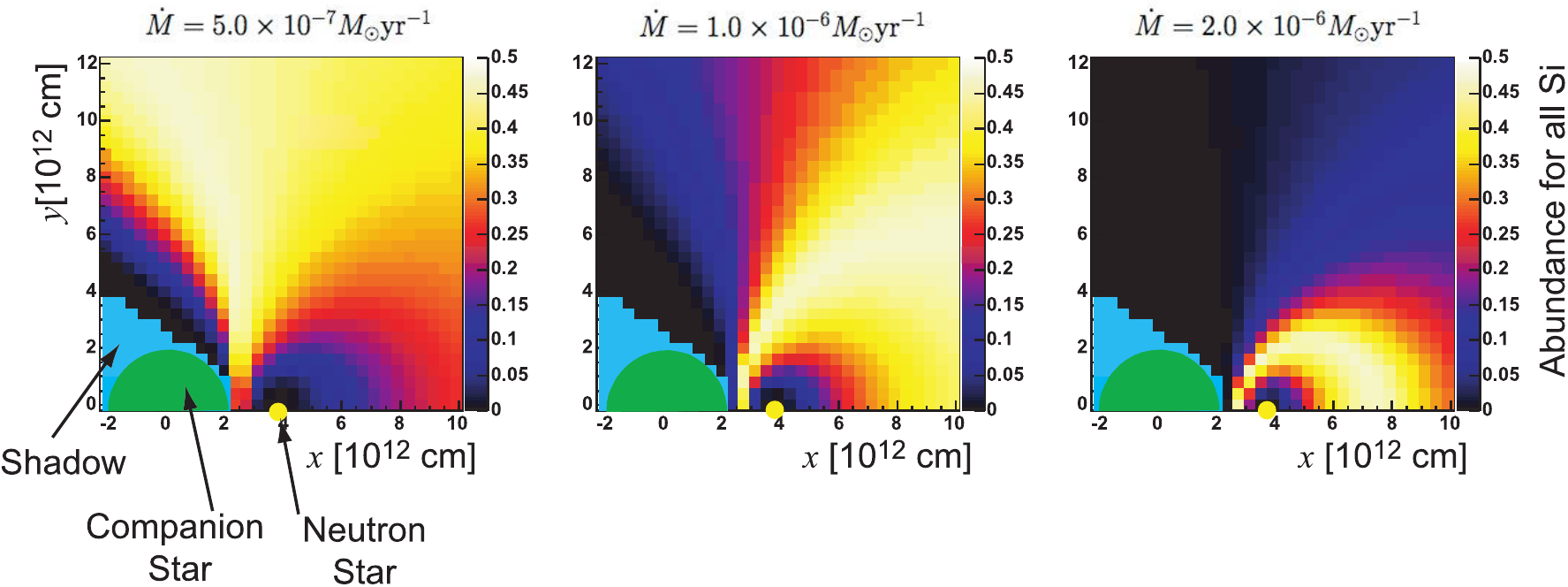}
\caption{The ion fraction maps of Si XIV (H-like) in the Vela X-1 system for different mass-loss rates $\dot{M}$ of the B-type donor star \citep{Watanabe:2006}.}
\label{fig:nsf/advanced_modeling/velax1_ionmap}
\end{center}
\end{figure}

\begin{figure}[htbp]
\begin{center}
\includegraphics[width=0.6\hsize]{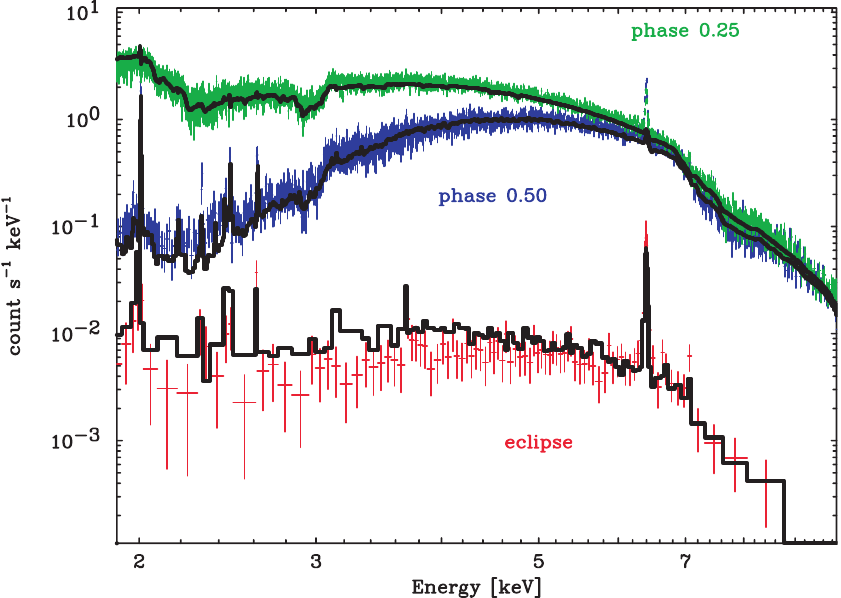}
\caption{Comparison of the simulated spectra of Vela X-1 with the observed data for three different orbital phases \citep{Watanabe:2006}. The normalization of the simulation models is fixed between the three phase ranges, assuming that the average X-ray luminosity does not change.}
\label{fig:nsf/advanced_modeling/velax1_spec}
\end{center}
\end{figure}

Photoionized stellar winds in high-mass X-ray binaries (HMXBs) provide ideal laboratories for study of physics of photoionized plasmas (see HMXB chapter).  \citet{Watanabe:2006} analyzed {\it Chandra} grating spectra of Vela X-1 and modeled the spectra by using Monte Carlo simulations. Figure~\ref{fig:nsf/advanced_modeling/velax1_ionmap} shows ion distributions of Si XIV in the binary system, which were calculated by using XSTAR, assuming a standard stellar wind model by \citet{CAK:1975}. Then, using these ion distributions, they performed the Monte Carlo simulation that carefully treats photoionization and photoexcitation, as well as Compton scattering. In this framework, resonance scattering is naturally included in a process of photoexcitation followed by a radiative decay. As shown in Figure~\ref{fig:nsf/advanced_modeling/velax1_spec}, the simulation reproduced the observed data with a mass-loss rate of $1.5\times 10^{-6}\ M_\odot \ \mathrm{yr}^{-1}$. This approach, using detailed Monte Carlo simulations, will be applicable to a wide variety of astrophysical objects including AGN outflows \citep[e.g.][]{Evans:2010,Detmers:2011,Tombesi:2010}.

\section{Radiative Transfer (including Compton Scattering)\label{sec:nsf/radiative_transfer}}

\subsection{Introduction}\label{subsec:nsf/radiative_transfer/introduction}

The unparalleled high-quality data obtained with {\it ASTRO-H} will require accurate and precise astrophysical interpretation. In addition to the detailed treatment of the emission processes described in the previous section (\S\ref{sec:nsf/advanced_modeling}), the modeling of electromagnetic radiation from astrophysical objects can involve solving problems of radiative transfer, since the emitted radiation may be altered by absorption as well as scattering.
A variety of methods for the radiative transfer problems have been developed; the selection of approaches depends on physical situations of the problem, required accuracy, and/or convenience of astrophysicists (and sometimes on their preference).

The optical thickness of astrophysical system is the most important physical quantity that characterizes the problem. If the system is optically thin, the problem becomes straightforward. For optically thick systems, we usually describe the problem by a partial differential equation where a continuous approximation is introduced, and then solve it analytically or numerically. Particular difficulties lie in the intermediate situation, namely, an optical depth about 0.5--10. Since in this case individual reprocessing of photons by matter has significant effects on the resultant emission, we have to treat the discontinuity of the photon processes, competing processes, and multiple interactions. Moreover, the geometry of the system, which can be complicated, should be taken into account.

\subsection{Monte Carlo Approach}\label{subsec:nsf/radiative_transfer/monte_carlo}

One of the standard approaches to the problems of radiative transfer is applying the transfer theory, which is based on the concept of a ``ray'' as a good approximation of radiation \citep{Rybicki:radiative_process}.
Introducing a specific intensity $I_\nu$ as the radiative energy propagating in a certain direction within a unit solid angle crossing unit area per unit time per unit frequency, the radiative transfer equation is formalized as
\begin{equation}
\frac{dI_\nu}{ds} = -\alpha_\nu I_\nu + j_\nu,
\end{equation}
where $ds$\ is a small length along the ray, $\alpha_\nu$ is the absorption coefficient, and $j_\nu$ is the emission coefficient. This equation has a more general form that can treat time dependence and multi-dimensional space coordinates.

For the radiative transfer of X-rays, however, it is more useful in many cases to treat the radiation as photons.
In this approach, the radiation can be calculated by a full photon tracking simulation based on a Monte Carlo method -- aka, a ``Monte Carlo simulation.''  This method provides a natural description of the problems particularly for the systems with intermediate optical depths, since the discrete nature of the photon processes is automatically involved, and the complexities of the photon interactions mentioned above can be easily handled.

\begin{figure}[htbp]
\begin{center}
\includegraphics[width=8.5cm]{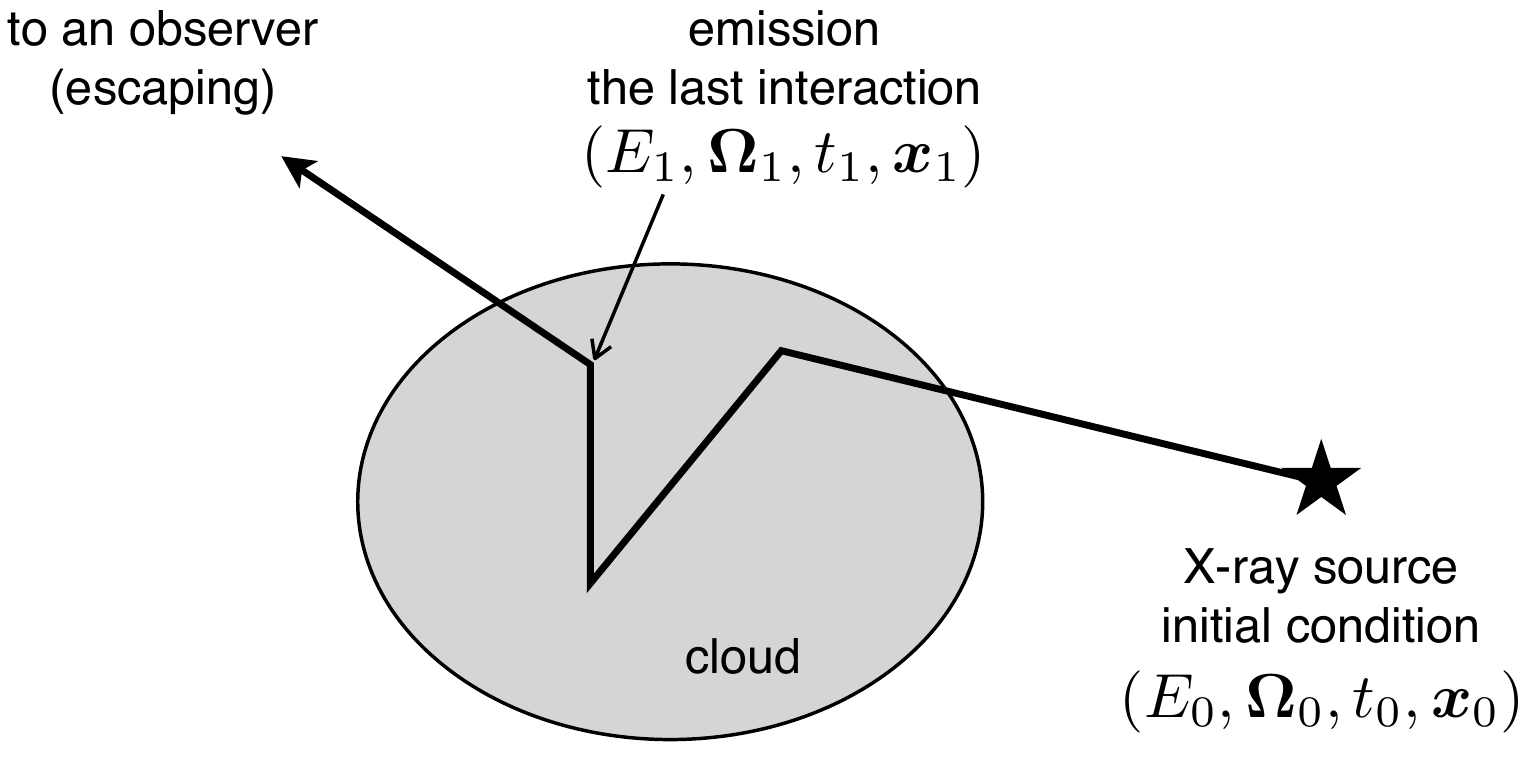}
\caption{A schematic procedure of the Monte Carlo simulation \citep{Odaka:PHD}. In one tracking trial, the photon is generated at the X-ray source, and then it is tracked until escaping from the system. The final interaction point can be regarded as an emission by an observer. The initial and final points should be recorded for a following analysis of the simulation result.}
\label{fig:nsf/radiative_transfer/mc_concept}
\end{center}
\end{figure}

The simulation calculates the propagation and interactions of the tracked photon with matter.
These are performed in two steps as shown in Figure~\ref{fig:nsf/radiative_transfer/mc_concept}.
First, after a photon starts from a certain initial point, the next interaction position is sampled from an exponential distribution characterized by a mean free path, 
\begin{equation}
l = \frac{1}{\sum n_i \sigma_i},
\end{equation}
where $n_i$ and $\sigma_i$ are the number density of a target and the cross section of a process labeled by $i$, respectively. A process that occurs at the determined position is selected from among all competing processes by a probability proportional to $n_i\sigma_i$.  Second, when the process occurs at the interaction position, the photon is absorbed or scattered.  If it is absorbed, one or more photons can be reemitted through a certain process such as radiative decay.  If it is scattered, the energy and direction are changed according to a differential cross section.  While tracking, these two steps are applied repeatedly until the photon is absorbed or escapes from the system.  The final interaction point before the escaping is recorded as an emission seen by an observer.

\subsection{X-ray Reflection and Compton Shoulder}\label{subsec:nsf/radiative_transfer/reflection}

When matter is illuminated by X-rays, we can observe X-rays reprocessed via photoelectric absorption followed by fluorescence, or Compton scattering. This phenomenon is called X-ray reflection. The X-ray reprocessing in cold matter provides us with an important probe of the structure, dynamics and chemical composition of the irradiated cold matter such as molecular clouds, dense gases in X-ray binaries, or torus-like structures around AGN. The reprocessed emissions also bring us fruitful information about the X-ray sources---black holes, neutron stars, or other energetic objects. This information becomes indispensable when we need to know the past activity of the illuminators by observing the delayed light via the reflector.
If the distance between the illuminator and the reflector is small, the illuminated matter gets ionized and spectral modeling becomes more complicated. The reflection from ionized matter has been modeled in a context of reflected X-rays from accretion disks around black holes \citep{Ross:2005, Garcia:2010}.

\begin{figure}
\begin{center}
\includegraphics[width=0.5\hsize]{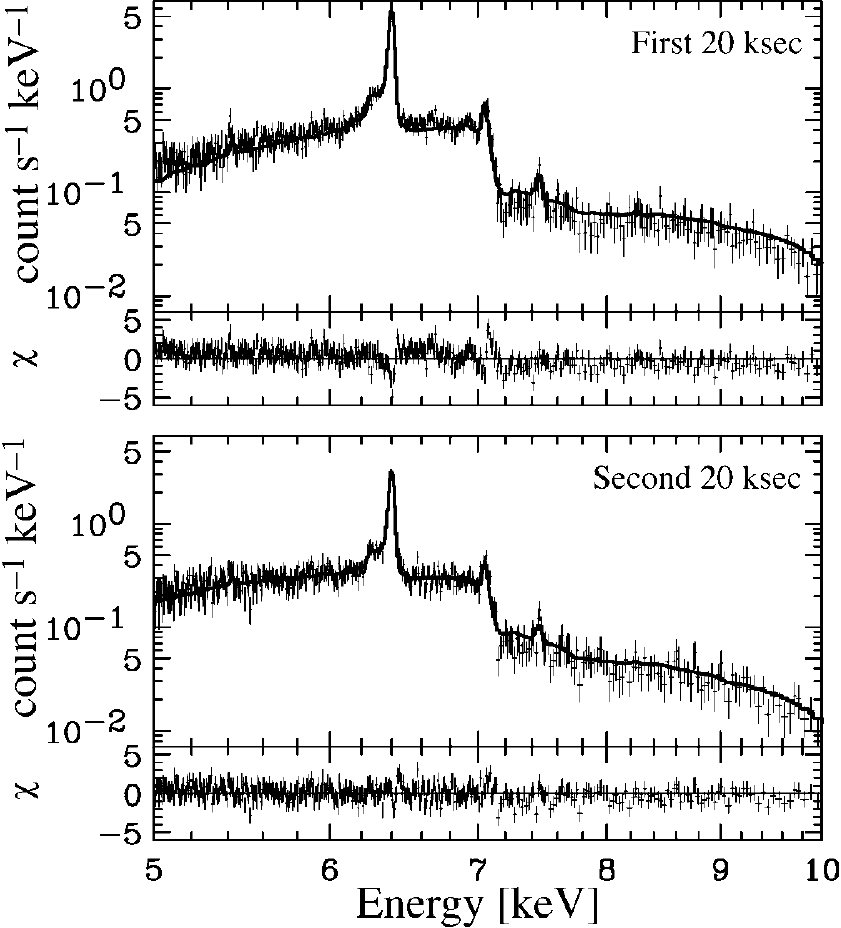}
\end{center}
\caption{Spectra of GX 301-2 from 5 to 10 keV obtained with {\it Chandra} HETG for the first and second halves of the observation \citep{Watanabe:2003}. The solid lines show the Monte Carlo models inferred from the data. In addition to the iron Ka emission line profile, the continuum shape is also successfully reproduced with the parameters inferred from the Compton shoulder profile. The values for $N_\mathrm{H}$ are $12.0\times 10^{23}\ \mathrm{cm}^{-2}$ and $8.5\times 10^{23}\ \mathrm{cm}^{-2}$ for the first and second halves, respectively.}
\label{fig:gx301}
\end{figure}

The reflection spectrum features a wide-band scattered continuum that has a hump at $\sim$20 keV (known as the ``Compton hump'') as well as fluorescent lines from the abundant heavy elements. In addition, if the reflecting cloud is not optically thin, scattering of fluorescent line photons in the cloud forms a low-energy tail structure at the corresponding fluorescent lines. This spectral feature, called a Compton shoulder, contains information about the scattering medium. Figure~\ref{fig:gx301} shows the spectrum of the high-mass X-ray binary GX 301-2 obtained with {\it Chandra} HETG, clearly displaying Compton shoulders around 6.2--6.4 keV of iron K$\alpha$ lines. By comparing the observed spectra and results of a detailed Monte Carlo simulation, \citet{Watanabe:2003} constrained physical conditions of the scattering medium in the binary system including the electron temperature, column density, and the metal abundance.

Since the X-ray reflection is widely seen in the Universe, it has been important to develop a calculation method of the spectrum and morphology. \citet{Murakami:2001} developed a three-dimensional model of giant molecular cloud Sagittarius B2 (Sgr B2), and compared its morphology with an image of the iron line obtained with {\it Chandra}. Sgr B2 is thought to be reflecting X-rays originated from a past outburst of the supermassive black hole Sgr A* at the Galactic center (GC). Hamaguchi et al.\ applied a similar approach to an X-ray reflection nebula in the $\eta$ Carinae complex to obtain its spectrum and time variability of the emission. X-ray emission from the supermassive star, $\eta$ Carinae, is reflected, or absorbed and reemitted, at the surrounding bipolar nebula called the Homunculus nebula. The nebula glows in X-rays and can be recognized in high-resolution images with {\it Chandra} when strong X-ray emission from the central binary system plunges at around its periastron passage for a couple of months (Corcoran et al. 2004). In a series of {\it Chandra} observations during the X-ray minimum phase in 2009, the glow gradually faded as expected in a simulation of X-ray reflection from the central binary system at the Homunculus nebula, as shown in Figure~\ref{fig:nsf/radiative_transfer/eta_carinae_hn}. The observed spectra showed a broad line at ~6.5 keV with a very high equivalent width of ~2 keV.  The line can be reproduced by iron fluorescence plus Compton down scatter of the Helium-like iron line in the incident spectra at the nebula.

\begin{floatingfigure}[r]{3.5in}
\includegraphics[width=3.2in]{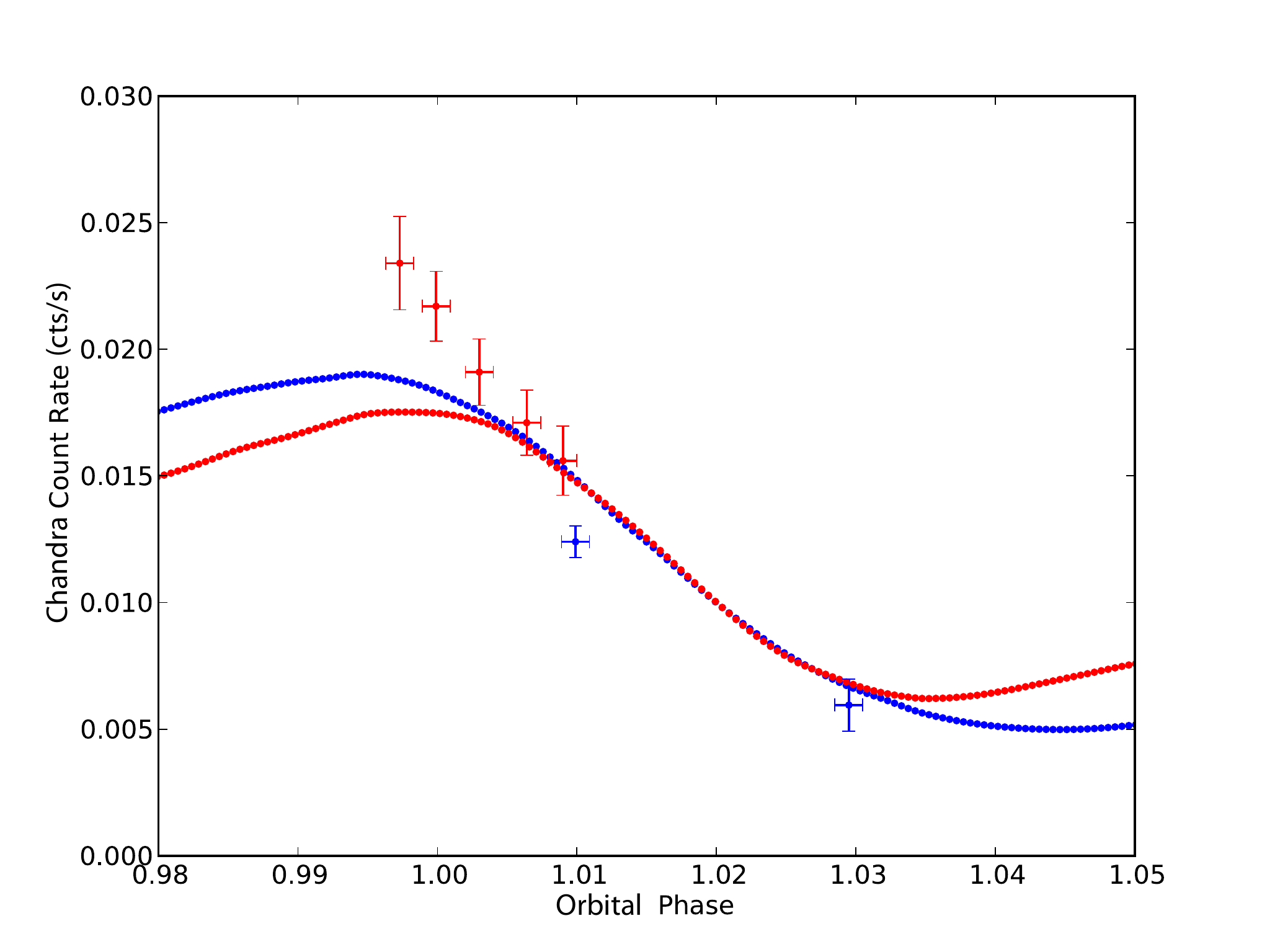}
\caption{
X-ray reflection light curve of the HN (blue points: 2003, red points: 2009) and 
the best-fit of the light curve to a time-delay reflection model using the observed {\it RXTE}  
light curve of the central binary system (blue and red solid lines) (Hamaguchi). The horizontal 
axis shows the orbital phase of the central binary system.}
\label{fig:nsf/radiative_transfer/eta_carinae_hn}
\end{floatingfigure}

In the {\it ASTRO-H} era, we will obtain more detailed information about the X-ray reflection.
The microcalorimeter will measure fine spectral structures such as the Compton shoulder, absorption edges, and line shift and broadening due to the dynamical motion including turbulence. Moreover, imaging spectroscopy of hard X-rays above 10 keV enables probing of the deep regions in illuminated clouds owing to their large penetrating power compared with that of soft X-rays.  Modeling the hard X-rays will also require  accuracy since the hard X-ray photons may experience multiple scattering in the dense clouds, introducing complexity to the radiative transfer.

To handle this problem, a Monte Carlo approach that includes accurate physical processes and realistic geometry is the most suitable though it is sometimes time-consuming. Several authors have been adopted Monte Carlo simulations for modeling the X-ray reflection \citep{Leahy:1993, Sunyaev:1998, Fromerth:2001}.
\citet{Odaka:PHD} also developed a multi-purpose framework (called MONACO) to calculate X-ray reprocessing based on a Monte Carlo simulation. This calculation includes detailed treatment of photoelectric absorption and scattering by electrons bound to hydrogen molecules and helium atoms \citep{Sunyaev:1996}.
The scattering by a bound electron makes a difference in the spectral shape of the Compton shoulder from Compton scattering by free electrons---an effect that will often be evident in the SXS spectrum.

\begin{figure}[htbp]
\begin{center}
\includegraphics[width=0.9\hsize]{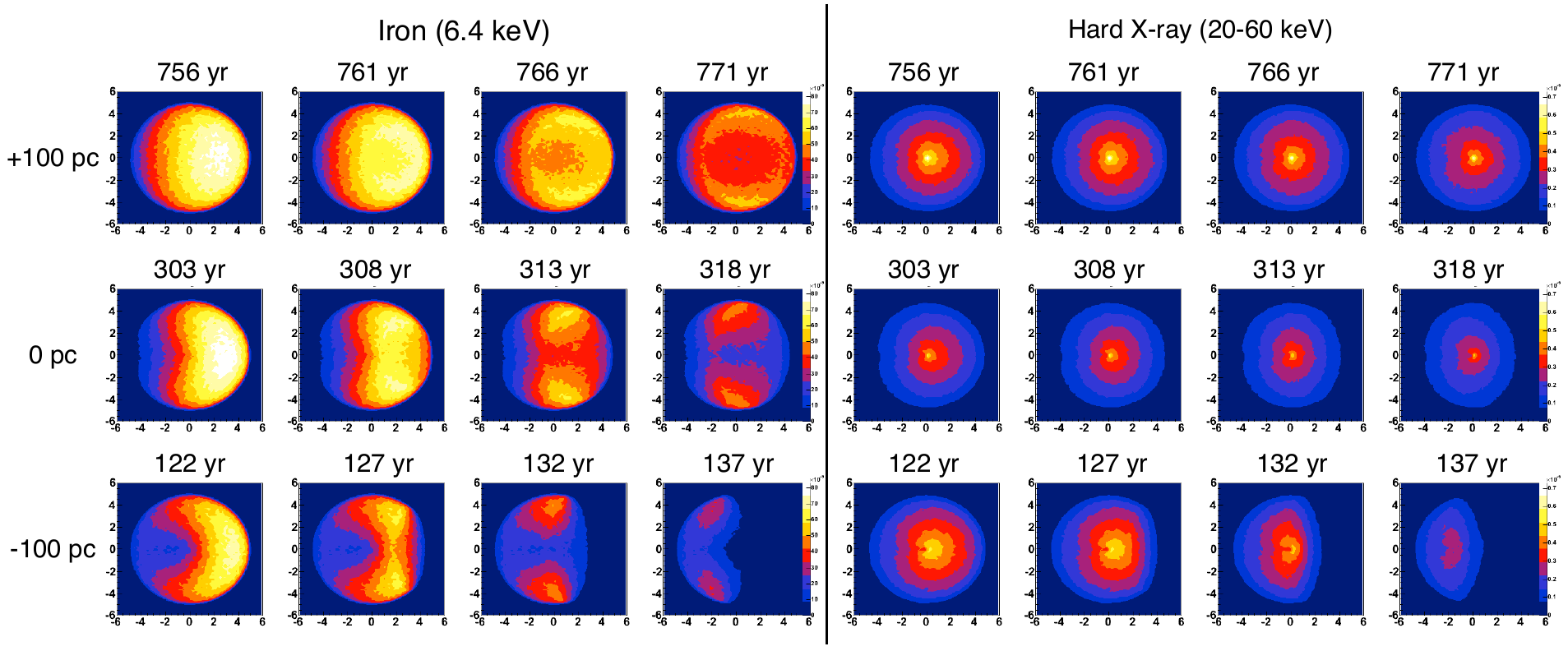}
\includegraphics[width=0.7\hsize]{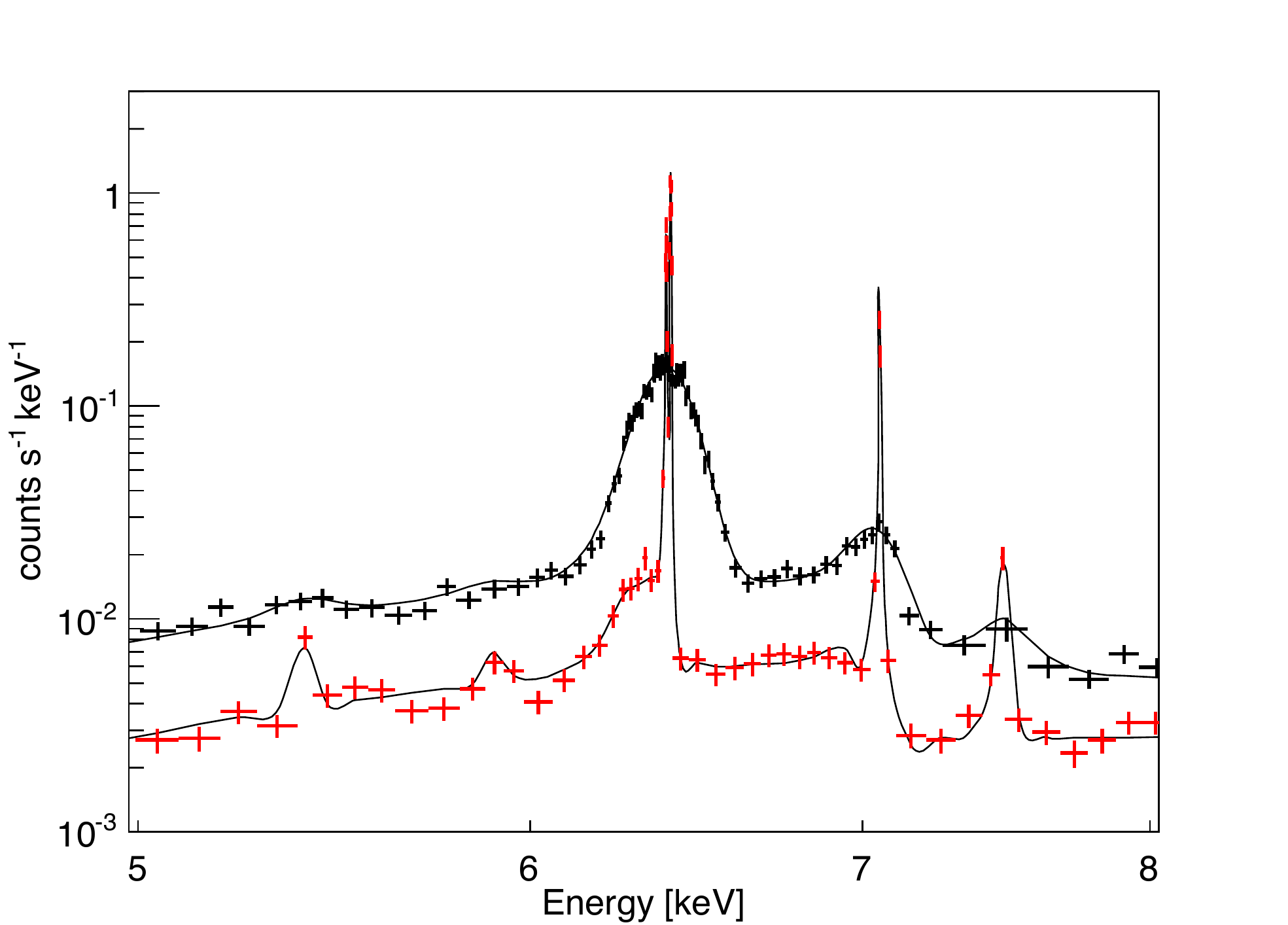}
\caption{
Calculated morphologies (top) and spectrum (bottom) of the X-ray reflection from Sgr B2 
\citep{Odaka:2011}. The top panel shows the time evolution of the morphology for different 
energy ranges (left: iron 6.4 keV line; right: 20--60 keV)  from the brightest moment at intervals 
of five years. The observation time is marked at the top of each image and the cloud position 
along the line of sight is marked at the left of each row ($t=0$ is the moment of the end of the 
outburst). The bottom panel shows observation simulations of Sgr B2 with {\it ASTRO-H} SXS (red 
points) and {\it Suzaku} XIS 0+3 (black points) for an exposure time of 200 ks. The flux is assumed 
to be a value in the fading phase, namely 15 years after the brightest moment.
}
\label{fig:nsf/radiative_transfer/sgrb2}
\end{center}
\end{figure}

By using the MONACO framework, \citet{Odaka:2011} calculated the X-ray reflection from Sgr B2 for several different models of the cloud structure, and predicted high-resolution spectra and hard X-ray morphologies, which can be observed by {\it ASTRO-H} as shown in Figure~\ref{fig:nsf/radiative_transfer/sgrb2}. Interestingly, the morphology of scattered hard X-rays above 20 keV is significantly different from that of iron fluorescence due to their large penetrating power. High-resolution spectra provide quantitative evaluation of the iron line including its Compton shoulder, constraining the mass and  chemical composition of the cloud as well as the luminosity of the illuminating source. Such careful modeling will be necessary to extract physical parameters of the cloud itself and the central black hole from the high-quality data. A large-scale view of X-ray reflection from molecular clouds in the GC region will reveal the structure of the GC molecular zone and the history of the activity of the central black hole in our Galaxy.

X-ray reflections in the vicinity of AGN contains important information on the structure and physical conditions of the environment around supermassive black holes ({\it See AGN reflection chapter of this White Paper}). Since the reflector, which is commonly referred to as an AGN torus, can be Compton-thick, Monte Carlo simulations are effective for accurate modeling that allows us to compare high-precision data obtained with the SXS and HXI. A set of numerical data tables called MYTorus \citep{Yaqoob:2012}, which is generated by Monte Carlo simulations, is useful for detailed modeling of obscured AGN.

\subsection{Photon Reprocessing in X-ray Emitting Plasmas}\label{subsec:nsf/radiative_transfer/emissions}

The reprocessing of photons can also have an impact on a spectrum emerging from an X-ray emitting plasma. Photons generated in the plasma can interact with electrons, atoms, or ions in the same plasma, altering the source function of the spectrum into other forms. If the assumption that the plasma is optically thin breaks, we should consider effects of radiative transfer. In this case, the Monte Carlo simulation allows us to construct precise models or to evaluate systematic uncertainties due to simplifications or assumptions of other fast models.

\subsubsection*{Photoionized Plasmas}

\begin{figure}[htbp]
\begin{center}
\includegraphics[width=8.0cm]{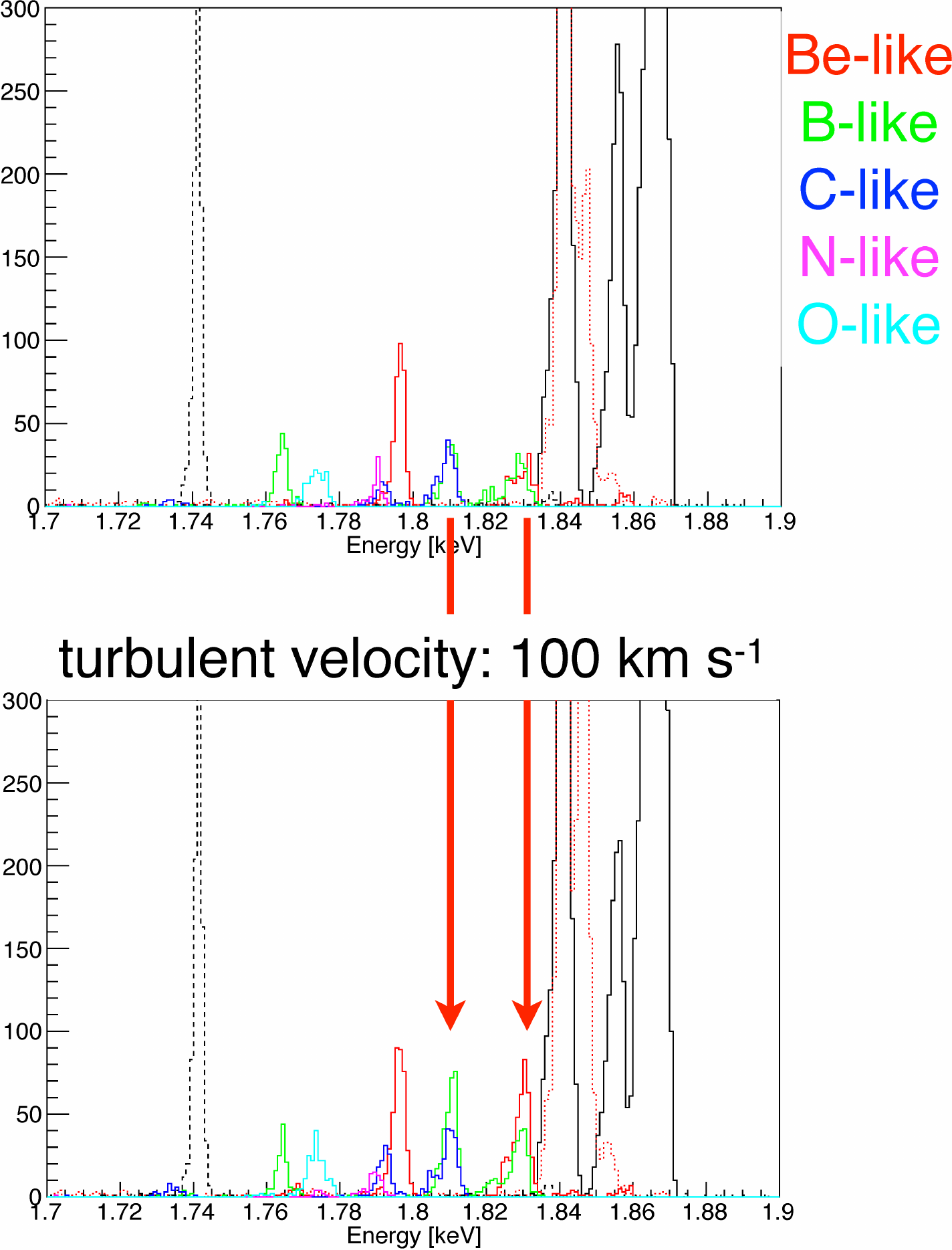}
\caption{The two spectra show difference in line emissions due to turbulent velocity \citep{Odaka:PHD}. This is the first attempt to reproduce Si-K complex of Vela X-1 spectrum obtained with {\it Chandra} though model parameters are not optimized yet. Contributions of each Si ions to the Si K-line complex are obtained by the simulation: Si XIV (black dotted), Si XIII (black solid), Si XII (red dotted), Si XI (red solid), Si X (green), Si IX (blue), Si VIII (magenta), Si VII (cyan), and Si I--VI (black dashed). Photon counts are represented in arbitrary units.}
\label{fig:nsf/radiative_transfer/si_complex}
\end{center}
\end{figure}

As discussed in \S\ref{subsec:nsf/advanced_modeling/photoionized_plasma}, radiative transfer plays a significant role in generating a spectrum of a photoionized plasma. For a photoionized plasma with complicated structure (e.g.\ stellar winds), the Monte Carlo simulation would be the most suitable approach. In Figure~\ref{fig:nsf/radiative_transfer/si_complex}, we show an effect of turbulence on the spectrum of Vela X-1 calculated by the simulation. Turbulence suppresses resonance-Auger destruction\citep{Ross:1996, Leidahl:2005}, resulting in enhancement of resonance lines.

\subsubsection*{Resonance Scattering}

Plasma diagnostics based on high-resolution spectroscopy with the SXS will open a new window to investigate physical properties (i.e.\ spatial distributions of velocity, temperature, or density) of hot X-ray emitting plasmas in the universe, such as clusters of galaxies, supernova remnants, and star forming regions. These plasmas are driven by collisional ionization and can be regarded as optically thin. (Collisional ionization equilibrium is not necessarily achieved in some cases such as supernova remnants or merging clusters). At energies of resonance lines, however, the optical depth of the plasma can be close to unity or larger, resulting in distortion of the line spectrum. This process affects, for example, shape of the line profile or intensity ratios between triplet lines of helium-like ions. These effects by resonance scattering are sensitive to a velocity field of both bulk and random motions of the gases. Therefore, they can be probes of gas dynamics including turbulence, which is believed to be one of the major forms of non-thermal energy in the Universe. The effects of resonance scattering on the plasma diagnostics can be simulated by careful treatment of radiative transfer (e.g. Monte Carlo simulations) \citep{Zhuravleva:2010}.

\subsubsection*{Comptonization}

Inverse Compton scattering by hot electrons or non-thermal electrons can be also considered as photon reprocessing in a plasma. Thus, Comptonization, which is a result of multiple scatterings, can be calculated by the photon tracking simulation. Though it is well known that the Kompaneets Equation or its extension gives the solution of Comptonization, Monte Carlo simulations become a suitable approach if the system has complicated geometry or is not thick enough for modeling by the differential equation. The simulation is also advantageous when the physical process is more complicated in the presence of a strong magnetic field, which introduces birefringence to photons and cyclotron resonance effects.


\section{Nuclear Lines, Annihilation Emission, and Solar Particle Acceleration}

\subsection{Introduction}

Spectral line features, primarily resulting from nuclear transitions, are present in the hard X-ray 
and $\gamma$-ray bands and detecting these will be the goal of many  \astroh/HXI and SGD 
observations.  In particular, the SGD will allow us to search for line emission in the soft 
$\gamma$-ray band from $\sim 80$ keV up to $\sim 600$ keV with unprecedented sensitivity. 
For example, {\it INTEGRAL}/SPI detected a strong electron-positron annihilation line at 511 keV from 
the direction of Galactic Center (GC) \citep{Jean03}. The SGD will enable us, for example, to 
search for not only long-term but also short-term variabilities from this promising source in the 
511~keV line thanks to its improved sensitivity. In addition, the radioactive decay of unstable 
isotopes synthesized during both supernova (SN) and nova explosions produces line features in 
the hard X-ray and $\gamma$-ray bands. In Table~1, we tabulate radioisotopes which emit line 
features in the energy ranges covered by the HXI and SGD \citep{DiehlTimmes98}.

Electron orbiting in a strong magnetic field have quantized energies, as do the X-rays that can be absorbed by these electrons.  These absorption features, known as cyclotron resonance scattering, have been observed in the hard X-ray spectra of some of neutron star X-ray binaries \citep[e.g.,][]{Mihara95, Makishima99}, and they can be investigated by \astroh/HXI. In addition, it is expected that solar flare neutrons are detectable by SGD. In the following, we describe the expected scientific results as well as potential \astroh/HXI and SGD targets which should show line features in the hard X-ray and $\gamma$-ray bands.

\subsection{Nuclear lines}
\subsubsection*{Scientific Context}
\begin{figure}[htbp]
\begin{center}
\includegraphics[width=0.5\hsize]{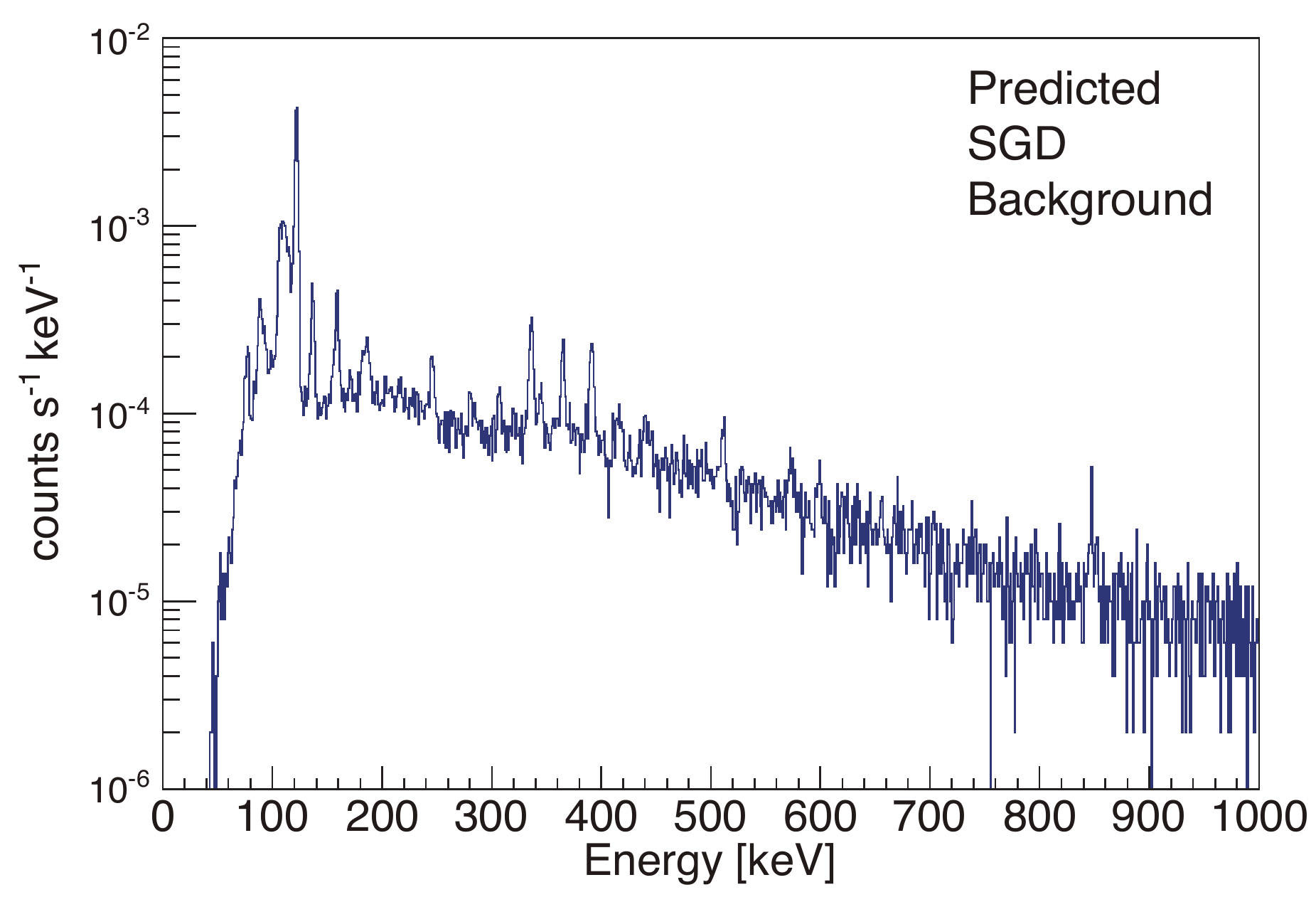}
\caption{Pre-launch prediction of {\it ASTRO-H} SGD background.\label{sgd_bgd}}
\end{center}
\end{figure}

Type Ia SNe, thermonuclear runaway explosions of white dwarfs in binary systems, synthesize huge amounts of $^{56}$Ni during the explosion itself \citep[e.g.,][]{Nomoto84}. $^{56}$Ni decays into $^{56}$Co via electron capture with a half-life of $\sim$6.0 days, producing 158 keV line emission. Hence, detection of line feature at 158 keV provides us with direct evidence of nuclear synthesis during SN explosions. So far the 158 keV line has not  been detected; upper limits have been obtained by \integral/SPI for nearby SNe Ia such as SN2011fe at a distance of $\sim$6.4 Mpc \citep{Isern11}. Measurement of the $^{56}$Ni line flux allows us to measure the total amount of $^{56}$Ni created during the supernova explosion, which is only indirectly estimated from light curve and absolute luminosity in optical band. For example, based on the \integral/SPI upper limit of the line flux of 2011fe, the amount of $^{56}$Ni produced is estimated to be $\lesssim 1.0 M_{\odot}$ \citep{Maeda12}.  Our efficiency will be affected by the SGD background; Figure~\ref{sgd_bgd}\ shows the latest simulated rates. 

$^{44}$Ti is also a detectable radioactive isotopes synthesized by core-collapse SNe. It is believed to power optical and infrared emissions from SN remnants after the first several years, emitting 67.9 and 78.4 keV lines during the decay chain of $^{44}$Ti $\rightarrow$ $^{44}$Sc$\rightarrow$$^{44}$Ca with a characteristic time of $\sim$85 yrs. Previous observations have already detected both of these lines from Cassiopeia~A and SN 1987A, and the synthesized mass of $^{44}$Ti was derived as $\sim \left( 1-3 \right) \times 10^{-4} M_{\odot}$ \citep[e.g.,][]{Vink01,Grebenev12}\ in each.

Several radioactive isotopes are synthesized also by classical nova explosions, the thermonuclear reaction on top of an accreting white dwarf in binary system. The nova's hydrogen burning phase should synthesize isotopes such as $^7$Be and $^{22}$Na \citep[e.g.,][]{Hernanz06}. $^7$Be (lifetime of 77 days) decays into an excited state of $^7$Li and releases a 478 keV photon when it de-excites. Although the photon energy is within the usable energy range of SGD, the total luminosity is estimated to be faint based on theoretical calculations \citep{Hernanz06}. $^{22}$Na also emits $\gamma$-ray photons with an energy of 1275 keV when it decays into $^{22}$Ne and de-excites, but this photon energy is above the SGD upper threshold of $\sim$600 keV. Given these difficulties, we do not consider here the detection possibility of line features from novae.

\subsection{Electron-positron annihilation emission}

\subsubsection*{Scientific Context}
An electron-positron pair annihilates into two 511 keV $\gamma$-rays. Since the first detection 
of the 511 keV line emission from the direction of the Galactic Center (GC) in 1973, the 
presence has been confirmed by subsequent numerous observations 
\citep[e.g.,][]{Jean03}. The 511 keV sky map obtained by most recent \integral/SPI 
observation revealed that the line emission is quite strong for the Galactic bulge direction 
centered on the GC and also showed symmetry with the spatial extension of $\sim 8^{\circ}$ 
\citep{Knodlseder05}. On the other hand, the emission from the Galactic disk is weak, 
suggesting that the bulge-to-disk ratio is very large. This is quite different from, for example, 
the MeV/GeV Galactic diffuse emission revealed by \fermi/LAT. The line flux of the bulge 
component is measured to be $\sim 1.0 \times 10^{-3}$ photons cm$^{-2}$ s$^{-1}$, but the 
origin of this strong bulge component is still unclear. On the other hand, positrons responsible 
for the disk component are thought to be produced by $\beta^+$\ decay of $^{56}$Ni, $^{44}$Ti 
and $^{26}$Al synthesized by SNe.

Although {\it INTEGRAL}/SPI has not yet discovered any point sources of the 511 keV line, in the past 
some papers claimed detection of transient emission of 511 keV line from a point source. 
Detection and variability of 511 keV line from 1E~1740.7-2942 (known as Great Annihilator) 
were reported in the literature \citep{Bouchet91}. \citet{Goldwurm92} also observed a 
significant 511 keV feature from Nova Musca. Radioactive decay of $^{22}$Na synthesized 
during nova explosions should produce a large amount of positrons through the $\beta^+$ 
decay process. Theoretical estimation shows that a strong 511 keV line with flux of 
$\sim 1.0 \times 10^{-4}$ photons cm$^{-2}$ s$^{-1}$  is emitted when assuming 
a distance of 10.5 kpc \citep{Suzuki10}, although this assumed a large amount of ejecta mass 
of $\sim 10^{-3} M_{\odot}$, which is much larger than $\sim 10^{-5} M_{\odot}$ derived from 
previous observations and theoretical calculations.

Recent \fermi/LAT detection of MeV/GeV emission from some of novae \citep[e.g.,][]{Abdo10} provide another channel of producing 511 keV line. In particular, the high-energy $\gamma$-ray emission detected by \fermi/LAT is thought to be produced by the decay of pions produced through $p-p$ interaction. If this is the case, positrons are also generated through decay of $\pi^+$. We should keep in mind, however, that the pion decay scenario of MeV/GeV emission from novae is not conclusive, and leptonic origin cannot be ruled out completely.

\subsection{Solar Neutrons}
The Sun sometimes emits flares whose radiation extends from radio to $\gamma$-ray bands. It is widely accepted that solar flare is caused by magnetic reconnection and the stored magnetic energy is converted into thermal and non-thermal particle velocities \citep[e.g.,][]{Tsuneta96}. Through ground-based and space-borne radio and X-ray observations of solar flares, our knowledge about electron acceleration has been greatly advanced. Ion acceleration, however, can be studied mainly by $\gamma$-ray observation through nuclear de-excitation lines and pion decay emission, but it has been difficult because of the low photon statistics. As a result, comparatively little is known about ion acceleration in solar flares.

Neutrons are also produced during flares through collision of accelerated ions with the solar atmosphere \citep[e.g.,][]{Chupp82}. Hence a measurement of solar neutrons allows us to study ion acceleration in solar flares. Solar neutron measurement has been performed by ground-based telescopes since 1970's, but as neutrons are significantly attenuated by the Earth's atmosphere, many assumptions and simulations are needed to infer the neutron spectrum above the atmosphere from observational data obtained by ground-based telescope.  Observations in space, of course, can directly and accurately measure the neutron spectrum using far fewer assumptions.

This neutron measurement will be performed with SGD combined with Monte Carlo simulation. Although the SGD is surrounded by a BGO scintillator to reject background, neutrons can penetrate the BGD shield and deposit part of the energy within Si and CdTe detectors. The incoming neutron spectrum will be derived by comparing the observed spectrum as a function of time with Monte Carlo simulation results of different normalizations and power-law indices. The neutron spectrum will provide us with data about accelerated protons above $\sim$30 MeV, while the proton spectrum above $\sim$300 MeV can be estimated from \fermi/LAT observations, if pion decay emission is strong enough. Thus, we can constrain the accelerated proton spectrum above $\sim$30 MeV and obtain information about total number and ratio of accelerated electrons and protons by combining data from other wavelengths.

\begin{table}[htdp]
\begin{center}
\caption{Lines from radioactive isotopes hopefully detected by \astroh/HXI and SGD (taken from Diehl).}
\label{default}
\begin{tabular}{ccccc}
\hline
Isotope & Life time & Decay chain & Energy [keV] & Source\\ 
\hline
$^{56}$Ni & 111 d & $^{56}$Ni$ \rightarrow$ $^{56}$Co$^{\ast}$ & 158, 812, 847, 1238 & SNe\\
$^{57}$Co & 390 d & $^{57}$Co$ \rightarrow$ $^{57}$Fe$^{\ast}$ & 122 & SNe\\
$^{44}$Ti & 89 yr & $^{44}$Ti $\rightarrow$ $^{44}$Sc$^{\ast}$ $\rightarrow$ $^{44}$Ca$^{\ast}$ + e$^{+}$ & 78, 68, 1157 & SNe\\
$^7$Be & 77 d & $^7$Be$ \rightarrow$ $^7$Li$^{\ast}$ & 478 & Novae\\
$^{22}$Na & 3.8 yr & $^{22}$Na$\rightarrow$ $^{22}$Ne$^{\ast}$ + e$^{+}$ & 1275 & Novae\\
\hline
\end{tabular}
\end{center}
\end{table}%



\appendix
\section{X-ray Spectral Diagnostics}

\subsection{General Principles}

X-ray emission spectra can be created by a range of high-energy processes, and can therefore encode information about those processes.  The table below lists the specific diagnostics relevant to the science identified by each {\it ASTRO-H} white paper team, but we begin with a brief review of the general possibilities.

{\bf Doppler effects:} Possibly the simplest diagnostic, using the position and width of an identified emission or absorption line relative to its natural position and width to infer the velocity and turbulence in the emitting or absorbing plasma. The primary challenges are (1) proper identification of the line, (2) ensuring that blending from other nearby lines is not affecting the measurement, and (3) obtaining an accurate laboratory wavelength for the line of interest.  Hydrogen-like and helium-like ion lines from $n=2\rightarrow 1$\ are well-known from both theory and measurement, but for most other lines in the X-ray the typical accuracy is only at the $50-100$~km~s$^{-1}$ level \citep{SB14}.

{\bf He-like triplets:}  The best-known of the X-ray line diagnostics, ratios of the $n=2\rightarrow 1$\ lines are functions of temperature, electron density, and ionization state \citep{Porquet10}.  Of course, in reality the system contains many more than three lines; there are four main transitions, and a much larger number of Li-like satellite features that can appear in the triplet `bandpass'.  

{\bf Resonance scattering:} Although it behaves in a similar manner as at longer-wavelength bandpasses, obtaining a sufficient column density of ions to observe resonance scattering can be difficult.  It has been detected in Fe~XVII in elliptical galaxies, as the Fe XVII 15.01\AA transition has a particular large oscillator strength, but the results are confused by ongoing difficulties in understanding the atomic physics of the Fe XVII ion \citep{dePlaa12}.

{\bf Compton shoulders:}  This feature is a low-energy tail on an emission line generated by Compton scattering of the line-energy photons. Since its spectral characteristics, including its shape, reflect the physical properties of the scattering medium, it potentially provides us with diagnostic tools of the density, electron temperature (or state of electrons), chemical/ion abundances, and geometry. Coupling of all these factors requires both precise modeling of the scattering processes and quantitative evaluation of spectral data.

{\bf Absorption edges:} Unlike the simple discontinuous jump found in the Henke tables, atomic absorption edges will have detailed structure created by solid state effects (X-ray Absorption Fine Structure, or XAFS), inner-shell excitation, excitation-autoionization effects, and so on.  This provides both an opportunity for new diagnostics if the observations, laboratory data, and instrumental calibration are adequate, or it can create confusion requiring the edge to simply be removed from the observations before modeling, as happened for years with the Si edge in many X-ray satellites.

{\bf General relativistic redshifts:} A strong gravitational field will shift the energy of emission lines; after integrating over the orbit, this has the effect of creating a broad redshifted tail in any emission line, although they are most typically seen in Fe K$\alpha$\ lines around AGN or Galactic black holes.  

{\bf Weak line features from nonthermal emission-dominated sources} Although nonthermal sources are often described as featureless power laws, there can exist weak line features due to underlying atomic processes induced by energetic photons, electrons, or ions. The detection of these features by the SXS will bring us new information on astrophysical situations in or around the high-energy sources.

{\bf Cyclotron line in strong B-fields:} In the presence of a strong magnetic field, electron motion perpendicular to the magnetic field is quantized into corresponding discrete energy levels (so-called `Landau levels'). Transitions between the Landau levels leads to absorption-like features in hard X-ray continuum from a magnetized neutron star. These features are also called cyclotron resonance scattering features (CRSFs) since the level transitions usually occur in the context of resonance scattering. Although astrophysical mechanism of generating CRSFs is not well understood, their spectral profiles, which will be obtained with the HXI, will contain important information about the magnetic field, plasma density and temperature, and geometry of the generation site.

{\bf K$\beta$\ diagnostics, satellite lines:} These are both categories of inner-shell transitions that occur in the presence of additional electron (or electrons) in the same shell or an outer shell.  This has the effect of moving the transition to somewhat lower energy compared to the same line created without the extra energy.  These lines may be weak, but they are useful ionization-state diagnostics since the ratio of the satellite line to the main line strongly depends upon the abundance of the two ions involved.

{\bf Non-equilibrium plasma:} In practice, electrons in astrophysical collisional plasmas seem to rapidly develop into Boltzmann distributions, although the ionization state of the plasma may take much longer to reach equilibrium.  As a result, collisional plasmas will typically either be under- or over-ionized to a degree dependent upon the age and density of the plasma.  

{\bf Charge exchange:} First detected in X-ray observations of comets, CX can be a strong process if the physical precursors of hot ions directly interacting with a neutral plasma can be created.  Although a new XSPEC model now exists, the atomic data needed for detailed spectral analysis is still lacking \citep{Smith12, Smith14}.

\subsection{Diagnostics by Science Area}

\subsubsection{Stars}

\begin{description}
\item[Protostars] {\it Time variation of the Doppler shifts of Fe K lines (6.7 keV, 6.4 keV)}: The spin and the radius of the central star of a protostar are determined with 6.7 keV line, and the dynamics and the spatial scale of the inner part of the accreting disk are determined with the fluorescent Fe-K line.
\item[Stellar flares] {\it The Doppler shifts of Fe K line (6.7 keV)}: Dynamic movement of materials during flares will be captured.
\item[Accretion in T Tauri star] {\it The density-sensitive lines}: Evidence for accreting plasma will be examined. 
\item[Massive stars, colliding wind binaries] {\it Equivalent width and shape of the Compton shoulder of Fe lines}:    
The geometry of wind-wind shock zones will be measured. It means that wind and orbital parameters of CWBs will be obtained.
\item[Diffuse X-rays in star forming regions] {\it Residuals in CCD spectra which could originate from unexpected lines indicating NEI conditions or charge exchange}:  X-ray emission mechanism(s) of the hot plasma which fills cavities in star forming regions.
\item[Star-planet interaction] {\it Ne line triplet}: Plasma density might be determined. Coronal emission might be distinguished from auroral charge-exchange emission, as has been observed in the polar regions of Jupiter.  \end{description}

\subsubsection{White Dwarfs}

\begin{description}
\item[Improved Mass Measurement] {\it G and R ratios of He-like Fe triplet lines}: Density and temperature structure of a post-shock region will be! extracted using a help of model calculation based on hydrostatics and plasma cooling functions. Precise determination of temperature structure will lead to WD mass estimation with higher accuracies.
\item[Coronal Disk Measurement] {\it Distorted Fe K emission line shape}: Rapid rotation of accretion disk of non-magnetic cataclysmic variables will show red-ward and blue-ward Doppler shift resulting in a distorted emission line profile. By measuring degree of distortion, location of disk coronae which is not well understood in cataclysmic variables will be estimated. 
\item[New Mass Measurement] {\it Gravitationally redshifted Fe K fluorescence line}: X-ray reflection at the WD surface will produce a redshifted Fe K fluorescence line, and precise measurement of energy shift will provide a new method of WD mass estimation that does not depend on orbital inclination.
\end{description}

\subsubsection{X-ray binaries, Magnetars, and Stellar-Mass Black Hole Systems}

Two diagnostics were identified for low-mass X-ray binaries, comprised of a compact object such as a neutron star or black hole and a low mass star (e.g. A, F, G, K or M).
\begin{description}
\item[Neutron star equation of state] {\it Gravitationally redshifted absorption line}: A significant detection of these absorption lines gives a clean measure of M/R.
\item[Wind]   {\it Absorption lines in winds}: photoionized wind model, geometry, such as the wind in emission from ADC LMXRB
\end{description}

Three were identified for high mass X-ray binaries, systems with a compact object and a massive O or B star. 

\begin{description}
\item[Wind characteristics] {\it  Spectrally and temporally resolve photoionization line profiles particularly of intermediate ionization stages of high-Z elements (Si, S, Fe) and of He-like ions}: Determine the wind characteristics of high mass stars, including density    temperature, turbulence, radial structure, and asymmetry (e.g., Vela X-1).
\item[Alfven shell] {\it Spin phase resolved fluorescent line profiles}: Investigation of kinematics of Alfven shell (e.g., GX 1+4, Her X-1) 
\item[Structure of wind] {\it Eclipse mapping of photoionization line strengths with high time resolution, e.g., in Vela X-1}:    
Determine the radial structure of the wind close to high mass stars 
\end{description}

Magnetars, neutron stars with extremely large magnetic fields, had two identified diagnostics.
\begin{description}
\item[Field strength measurement]   {\it Search for proton cyclotron resonance features of magnetars and magnetar-relatives like CCOs.}:
Direct evidence on the strong magnetic field and interior of magnetars (e.g., 4U 0142+61 and activated sources: see, Tiengo et al., Nature, 2013) 
\item[Age estimate] {\it Thermal plasma diagnostics of magnetar-associated supernova remnants}: Progenitors of magnetars via the abundance study and magnetar true age via measurements of the Doppler velocity of spectral lines. (e.g., Kes 73 and CTB 109)
\end{description}

Stellar mass black hole systems are a subclass of a X-ray binaries, but have unique science due to the nature of the compact object.

\begin{description}
\item[Wind] {\it  Detection and characterization of even very weak absorption lines}: A strong wind can alter the implied mass accretion rate, which is important to continuum measurements.
\item[Spin] {\it A forest of low-energy atomic emission lines, the Fe line, and the Compton back-scattering hump}: Strong constraint or measurement of black hole spin 
\item[Accretion Physics] {\it Energy-dependent polarization measurements}: To constrain spectral models and reveal the physical state of accreting matter. 
\end{description}

\subsubsection{Supernova remnants}

\begin{description}
 \item[Kinematics] {\it Doppler-shifted and broadened line structure of each element}: Determine the velocity distribution of each element, including expanding velocity, turbulence, anisotropy of ejecta expansion.
\item[Plasma Physics]  {\it Forbidden and resonance line ratio of Oxygen in shocks of old SNRs}: Determine the influence of charge exchange and resonance scattering effects in the plasma emission.
 \item[Abundances] {\it Line broadening in old SNR ejecta}: Determine the absolute abundance of old SNR ejecta like Vela shrapnel.  In the pure metal case, lines will broaden due to high ion temperature. 
\item[Shock Physics] {\it Satellite lines of iron K in over-ionized SNRs}: Determine how the shock exchanges thermal energy with interstellar medium.
\item[Pulsar Wind Nebulae] {\it Detecting lines from PWN dominated SNRs}: Determine thermal energy and abundance pattern to understand progenitors which made the neutron stars.
 \item[Reverse Shock] {\it Measurements of the Fe K$\beta$\ line}: Diagnostic of the reverse shock in young supernova remnants, such as Tycho
\end{description}

  
\subsubsection{Clusters of galaxies}
  
\begin{description}
\item[Cold fronts and mergers] {\it Doppler shift and broadening of the Fe K lines}: 
Measure turbulent and bulk motions in the intracluster medium (ICM) in merging clusters such as Coma and A3667, in order to quantify the spectrum of turbulence and the velocity field across cold fronts. The ultimate goal is to constrain the ICM microphysics.
\item[Gas dynamics, AGN feedback] {\it Doppler shift, broadening and resonant line scattering of the Fe-K lines}:Measure turbulent and bulk motions in the brightest cool cores  (e.g. Perseus, M87)  to constrain AGN feedback 
\item[Chemical composition] {\it Emission lines in collisional plasmas}: Measure metal abundances of various elements (C, N, O, Ne, Mg, Al, Si, S, Ar, Ca, Cr, Mn, Fe, and Ni) to constrain the metal enrichment histories in galaxies in clusters.
\item[WHIM] {\it Emission line strengths of O}:  Detect WHIM filaments
\item[Non-thermal pressure and mass] {\it Doppler shift and broadening of Fe K lines}: Quantify non-thermal pressure support due to bulk motion and! turbulence, understand cluster growth, energy dissipation in the ICM, cluster masses, and, ultimately, cosmological inferences based on cluster mass estimates 
\end{description}

\subsubsection{AGNs}
      
\begin{description}
\item[Reflection] {\it Shape of the Compton shoulder of fluorescent Fe-K line}: Geometry and optical depth of X-ray reflection form gas/dust surrounding the supermassive black hole.
\item[Winds] {\it Doppler shifts and strengths of Fe XXV and Fe XXVI lines}: Determine the physics and parameters of the AGN wind, including the total mass in the wind 
\end{description}

%


\subsubsection{High-z Universe}
    
\begin{description}    
\item[High-z chemical evolution] {\it Resonant absorption features in background GRB afterglow or blazar spectra}:   
Chemical evolution in the universe by identifying absorption element and measuring its ionization state due to intergalactic medium (WHIM). The expected temperature, ionization parameter (xi), and redshift are $10^{5Ð6}$\,K, $\sim 20$\,assuming CXB, and z=1Ð3, respectively. 
\item[GRB] {\it Emission Line features of Fe-K and other light elements (Mg, Si, S, Ar, and Ca)}: 
Redshift estimation of GRB; chemical evolution study using the GRB associated supernova ejecta; determination of the ejecta mass and velocity.
\item[GRB] {\it Detection and shape of radiative recombination edge}: GRB associated SN's ejecta mass and information on its progenitor by the measured edge energy and the optical depth. 
\end{description}


\clearpage
\begin{multicols}{2}
\end{multicols}

\end{document}